\documentclass[preprint]{aastex}
\usepackage{emulateapj5}
 
\newcommand{\zem}{{\ifmmode{z_{em}}\else{$z_{em}$}\fi}}
\newcommand{\zabs}{{\ifmmode{z_{abs}}\else{$z_{abs}$}\fi}}
\newcommand{\kms}{{\ifmmode{{\rm km~s}^{-1}}\else{km~s$^{-1}$}\fi}}
\newcommand{\delv}{{\ifmmode{\Delta v}\else{$\Delta v$}\fi}}
\newcommand{\cmm}{{\ifmmode{{\rm cm}^{-2}}\else{cm$^{-2}$}\fi}}
\newcommand{\cmmm}{{\ifmmode{{\rm cm}^{-3}}\else{cm$^{-3}$}\fi}}
\newcommand{\nhi}{{\ifmmode{N_{\rm H\;I}}\else{$N_{\rm H\;I}$}\fi}}
\newcommand{\nhe}{{\ifmmode{N_{\rm He\;II}}\else{$N_{\rm He\;II}$}\fi}}
\newcommand{\logn}{{\ifmmode{\log N}\else{$\log N$}\fi}}
\newcommand{\logz}{{\ifmmode{\log Z/Z_{\odot}}\else{$\log (Z/Z_{\odot})$}\fi}}
\newcommand{\lognhi}{{\ifmmode{\log\; N_{\rm H\; I}}\else{$\log\; N_{\rm H\; I}$}\fi}}
\newcommand{\lognhii}{{\ifmmode{\log\; N_{\rm H\; II}}\else{$\log\; N_{\rm H\; II}$}\fi}}
\newcommand{\lognciv}{{\ifmmode{\log\; N_{\rm C\; IV}}\else{$\log\; N_{\rm C\; IV}$}\fi}}
\def\lsim{\lower0.3em\hbox{$\,\buildrel <\over\sim\,$}}
\def\gsim{\lower0.3em\hbox{$\,\buildrel >\over\sim\,$}}

\newcounter{species} 
\def\ion#1#2{\setcounter{species}{#2}#1$\;${\scriptsize\Roman{species}}\relax}

\newcommand{\lya}{Ly$\alpha$}

\slugcomment{\today}
\shorttitle{Magellanic Bridge as a DLA System}
\shortauthors{Misawa et al.}

\begin{document}

\title{The Magellanic Bridge as a DLA System: Physical Properties of
Cold Gas toward PKS~0312$-$770\altaffilmark{1}}

\altaffiltext{1}{Based on observations obtained with the NASA/ESA
       Hubble Space Telescope, which is operated by the Space
       Telescope Science Institute (STScI) for the Association of
       Universities for Research in Astronomy, Inc., under NASA
       contract NAS5D26555.}

\author{Toru Misawa\altaffilmark{2,3},
        Jane C. Charlton\altaffilmark{2},
        Henry A. Kobulnicky\altaffilmark{4},
        Bart P. Wakker\altaffilmark{5}, and
        Joss Bland-Hawthorn\altaffilmark{6}}

\altaffiltext{2}{Department of Astronomy \& Astrophysics, The
  Pennsylvania State University, University Park, PA 16802}
\altaffiltext{3}{Cosmic Radiation Laboratory, RIKEN, 2-1 Hirosawa,
  Wako, Saitama 351-0198 Japan}
\altaffiltext{4}{Department of Physics \& Astronomy, University of
  Wyoming, Laramie, WY 82070}
\altaffiltext{5}{Department of Astronomy, University of
  Wisconsin-Madison, 475 North Charter Street, Madison, WI 53706}
\altaffiltext{6}{School of Physics, University of Sydney, NSW 2006,
  Australia}

\email{misawa, charlton@astro.psu.edu, chipk@uwyo.edu,
  wakker@astro.wisc.edu, jbh@aao.gov.au}

\begin{abstract}
We measure the physical properties of a local multi-component
absorption-line system at $V_{\odot}$ $\sim$200~\kms\ toward the
quasar PKS~0312$-$770 behind the Magellanic Bridge (MB) using Hubble
Space Telescope STIS spectroscopy in conjunction with photoionization
modeling.  At an impact parameter of $\sim$10~kpc from the Small
Magellanic Cloud (SMC), this sightline provides a unique opportunity
to probe the chemical properties and ionization structure in a nearby
absorption line system with a column density of \lognhi\ $\sim$20.2,
at the transition between Damped Lyman Alpha (DLA) and sub-DLA
systems.  We find that metallicity of $-1.0$ $<$ \logz\ $<$ $-0.5$ and
ionization parameter of $-6$ $<$ $\log U$ $<$ $-5$ for three
low-ionization components and $\log U$ $\sim$ $-2.6$ for one
high-ionization component.  One component at $V_{\odot}$ = 207~\kms\
shows an $\alpha$-element abundance $\log$(Si/H) $\sim$$-5.0$, making
it $\sim$0.2~dex more metal rich than both SMC \ion{H}{2} regions and
stars within the MB and the SMC.  The N/Si ratio in this component is
$\log$(N/Si) = $-0.3\pm0.1$, making it comparable to other N-poor
dwarf galaxies and $\sim$0.2~dex lower than \ion{H}{2} regions in the
SMC.  Another component at $V_{\odot}$ = 236~\kms\ shows a similar
Si/H ratio but has $\log$(N/Si) = $-1.0\pm0.2$, indicating a nitrogen
deficiency comparable to that seen in the most N-poor DLA systems.
These differences imply different chemical enrichment histories
between components along the same sightline.  Our results suggest
that, if these absorbers are representative some fraction of DLA
systems, then 1) DLA systems along single sight-lines do not
necessarily represent the global properties of the absorbing cloud,
and b) the chemical composition within a given DLA cloud may be
inhomogeneous.
\end{abstract}

\keywords{Magellanic Clouds -- galaxies: individual (Large Magellanic
Cloud, Small Magellanic Cloud) -- galaxies: abundances --- absorption
lines}

\section{Introduction}

A physical connection between the Large Magellanic Cloud (LMC) and the
Small Magellanic Cloud (SMC) was implied as early as 60 years ago
\citep{sha40}. A continuous \ion{H}{1} gas structure between the LMC
and the SMC, known as the Magellanic Bridge (MB), was first reported
by \citet{hin63}, and it is found to be at a distance of
$\sim$50--60~kpc \citep[e.g.,][]{har03}. There is substantial evidence
for a young stellar population in the MB, with stars as young as $\sim
20$~Myrs \citep{irw85,irw90,gro92,ham94,dem98} and with molecular
clouds detected \citep{leh02,mul03,miz06}.  These young populations
were probably formed locally, because they are not old enough to have
escaped from the next nearest star-forming region (i.e., the SMC)
based on their peculiar motions. On the other hand, the larger-scale
tidal structure that is connected to the MB, the Magellanic Stream,
contains only gas, with no evidence of a stellar population
\citep{guh98}. The physical connection of these structures are still
under discussion \citep{kal06,pia08}.

Although the formation mechanism of the MB is not understood in
detail, gravitational tidal interactions probably played a primary
role \citep[e.g.,][]{gar94,mur80}. Numerical simulations indicate that
the MB was created through a close interaction between the LMC and the
SMC about 0.2~Gyr ago \citep[e.g.,][]{gar96}. Thus, the MB is an ideal
target for studies of the influence of dynamical interactions on
star-forming activity. The MB also provides a unique opportunity to
study nearby star formation in a low metallicity environment.  The MB
stars have metallicities of only one-tenth that of normal Population I
Galactic stars (i.e., even smaller than the SMC stars by
$\sim$0.5~dex; \citealt{rol99}).

The abundance pattern of the MB has been studied through stellar
populations and gas clouds that are detected as absorption features in
the stellar spectra \citep[e.g.,][]{
dem98,leh01,leh08,leh02,mal01,wel01,li06,nis07,har07}.  Recently,
\citet{car08} found a possible metallicity gradient in the SMC toward
outer regions from $\sim$1 to $\sim$4 degree from its center.
However, stars and circumstellar regions could be biased because their
physical conditions are significantly influenced by nearby
stars. Analysis of interstellar gas should yield more representative
properties of the MB and its abundance pattern.

Quasar absorption lines are powerful tools to investigate the physical
conditions of absorbers located along sight-lines to quasars.  These
absorbers are detected in quasar spectra, regardless of the source
luminosity, which enables us to collect a homogeneous sample of
absorbing clouds from early cosmic epochs to the present. Although
there have been several attempts to trace the MB gas along the
sight-lines toward quasars, these were based on radio-spectra that
covered \ion{H}{1}~21-cm absorption \citep[e.g.,][]{kob99} or optical
spectra that covered the \ion{Ca}{2} absorption doublet
\citep{smo05}. On the other hand, most important ions (e.g.,
\ion{Mg}{1}, \ion{Mg}{2}, \ion{Si}{2}, \ion{Si}{4}, and \ion{C}{4})
have their resonance transitions in the rest-frame ultra-violet (UV)
region. Their observed wavelengths are still in the UV region for
low-redshift targets like the MB.

In this paper, we report on the general physical conditions and
chemical composition of the MB gas far from the location of stellar
populations, through quasar absorption line analysis. We chose the
sight-line toward a radio-loud quasar PKS~0312$-$770 at $z$ = 0.223
that passes through the MB (Figure~1) at a heliocentric distance of
$\sim$50~kpc (similar to the distance of the SMC). Because the quasar
is radio loud we are also able to detect neutral hydrogen gas as
\ion{H}{1}~21-cm emission/absorption features. \citet{kob99} reported
detections of such features with three components at a heliocentric
velocity of $V_{\odot}$ $\sim$200~\kms. Based on their fitting
results, they estimated a total neutral hydrogen column density of
$N_{HI}$ = 1.2$\times$10$^{20}$~\cmm\ and mean spin temperatures of
$\langle T_{S} \rangle$ = 22, 29, and 46 $K$ for the three components.
Although the measured spin temperatures are contaminated by warm gas
structures that are not related to the absorbing components, they are
similar to the spin temperatures of the LMC ($T_S$ = 30--40~$K$;
\citealt{meb97,dic94}) and the SMC ($T_S$ = 20--50~$K$;
\citealt{dic99}), and slightly smaller than that of M31 ($T_S$
$\sim$70~$K$; \citealt{bra92}).

Recently, \citet{smo05} detected \ion{Ca}{2}~K lines from the MB gas
in their optical, medium-resolution spectrum of PKS~0312$-$770.  Here
we report on detections of 30 additional metal absorption lines in an
{\it HST}/STIS UV high-resolution spectrum of the same quasar.  We
report the results of photoionization models in order to probe the
physical conditions in the MB along this sightline.

We describe the observations and data reduction in \S~2 and our
procedures for line profile fitting and photoionization modeling with
Cloudy \citep{fer98} in \S~3 and \S~4. We present our results in \S~5,
and discuss them in \S~6.

\section{Observations and Data Reduction}

A UV spectrum of the quasar, PKS~0312-770 ($m_{V}$ = 16.2), was taken
with the Hubble Space Telescope ({\it HST})/ Space Telescope Imaging
Spectrograph (STIS) (\citealt{kob00}), using two echelle gratings: the
E140M grating covering 1150\AA--1700\AA\ and the E230M grating setting
that covers 2130\AA--2985\AA, with 6--10~\kms\ velocity resolution.
The total exposure times are 37,908 sec for the E140M grating and 6060
sec for the E230M grating, which results in a spectrum with a S/N
ratio of 6--28 per pixel\footnote{It is typically $\sim$6 because only
a tiny range of wavelength around \lya\ emission line has higher S/N
ratio.} ($\Delta \lambda$ $\sim$0.015\AA) at $\lambda$ $<$1700\AA\ and
3--8 per pixel ($\Delta \lambda$ $\sim$0.04\AA) at $\lambda$
$>$2130\AA. A journal of the observations is shown in Table~1.

The data were processed with the standard {\it HST}/STIS pipeline,
{\sc CALSTIS} \citep{bro02}. Continuum fits were made with standard
techniques \citep{chu01} using the IRAF SFIT task.\footnote{IRAF is
distributed by the National Optical Astronomy Observatories, which are
operated by AURA, Inc., under contract to the National Science
Foundation.} The normalized spectrum, after rebinning every 0.15\AA\
for $\lambda$ $<$1700\AA\ and every 0.40\AA\ for $\lambda$ $>$2130\AA,
is presented in Figure~2 along with the 1$\sigma$ error spectrum.

The STIS spectrum covers various absorption lines detected at
$>5\sigma$, including (\lya, \ion{Mg}{1}, \ion{C}{1}, \ion{O}{1},
\ion{N}{1}, \ion{Mg}{2}, \ion{Mn}{2}, \ion{Fe}{2}, \ion{Si}{2},
\ion{Ni}{2}, \ion{S}{2}, \ion{C}{2}, \ion{Si}{3}, \ion{Si}{4},
\ion{C}{4}, and \ion{N}{5}) arising either from local absorbers in the
Milky Way, from high-velocity clouds, from the MB or from higher
redshift absorbers at $z$ = 0.1983, 0.2018, 0.2026, and 0.2029
\citep{gia05}. Absorption lines arising from the MB (that are our main
targets) are marked in Figure~2, and their rest-frame equivalent
widths or detection limits are listed in Table~2. We can effectively
separate absorption profiles of the MB from those of the Milky Way and
the high-velocity clouds, because their velocity separations are large
enough (the MB and the HVC are redshifted from the Milky Way by
$\sim$200 and 300 \kms, respectively). Only \lya\ is not separated
because it is hidden in the broad damping wing of the Milky Way
absorption. The total column density of this system (including
contributions from the Milky Way, the MB, and the HVC altogether) was
measured to be \logn~(\ion{H}{1}/[\cmm]) $\sim$20.85 \citep{gia05}, or
$\sim$20.87 \citep{leh08}.

As estimated below using photoionization modeling, the total
\ion{H}{1} column density of the MB toward the quasar PKS~0312-770 is
\logn~(\ion{H}{1}/[\cmm]) $\sim$20.2, just below the lower-limit of
the criterion for damped \lya\ system (DLA). The total rest-frame
equivalent width of \ion{Mg}{2} ($W_r$(2796) = 1.91\AA) is also large
enough for classification as a strong \ion{Mg}{2} system (whose
minimum equivalent width is $W_r$(2796) = 0.3\AA). Absorption systems,
classified as strong \ion{Mg}{2} systems, are usually associated with
bright ($L$ $>$ 0.05$L^{*}$) galaxies within 40$h^{-1}$ kpc (e.g.,
\citealt{ber91}). The ionization state of the system, in which we
detect very strong low-ionization transitions such as \ion{O}{1} and
\ion{Fe}{2} and which has either weak or undetected high-ionization
transitions (e.g., \ion{C}{4} and \ion{N}{5}), is very low, as is
often the case with DLA systems (e.g., \citealt{lu96}).  Thus, this
system in the MB is probably a local counterpart of DLA systems at
higher redshift. We also confirmed that all absorption features in
this system are clean without contamination by absorption lines of
other systems at higher redshift.

Visual inspection of the velocity plot of the MB system also suggests
that the physical and chemical conditions of this system are somewhat
similar to those of another absorber detected in the MB toward the
young stars, DI~1388 and DGIK~975 \citep{leh01,leh08,leh02}: (i)
having absorption lines with various ionization potentials from
\ion{O}{1} and \ion{C}{1} to \ion{C}{4} and \ion{Si}{4}, and (ii)
showing strong neutral transitions such as \ion{O}{1} and
\ion{N}{1}. \citet{leh01} derived several properties of the system
toward DI~1388: (a) dust depletion is moderate and its pattern is
similar to that of the Galactic halo, and (b) the metallicity is lower
than that in the solar neighborhood by 1.1~dex, and still lower by
0.5~dex than in the SMC. Using photoionization models, we estimate
these parameters toward the quasar, PKS~0312$-$770, separated by
$\sim$4.1 degree (i.e., $\sim$4.3~kpc at the distance of the SMC) from
DI~1388, so that we can study the variety of physical conditions in
the MB.

\section{Line Profile Fitting}

We determine the number of components needed to reproduce the observed
spectrum and measure their line parameters (i.e., column density,
Doppler parameter, and radial velocity) using Voigt profile
fitting. \citet{kob99} detected three \ion{H}{1}~21-cm absorption
components in their radio spectrum of PKS~0312-770 taken with the
Australia Telescope Compact Array (ATCA).  The \ion{H}{1}~21-cm
absorption lines are due to cool atomic hydrogen regions that probably
have similar physical conditions to DLA systems.

Following \citet{kob99}, we first used three \ion{H}{1} components to
fit other metal absorption profiles in our {\it HST}/STIS spectrum.
However, we found a three component fit insufficient because the line
widths of the three components presented in \citet{kob99} (FWHM = 9.6,
13.2, and 10.5~\kms, corresponding to $b({\rm H})$ = 5.8, 7.6, and
6.3~\kms) could not reproduce the wing profiles at both sides of the
metal absorption features.  For example, Figure~3 shows a comparison
between the observed and modeled \ion{O}{1}~$\lambda$1302 profiles.
We assumed that the Doppler parameters of oxygen and hydrogen were the
same for this purpose, i.e. that turbulence dominates $b$, since that
yields the maximum value for $b({\rm O})$.  If we increase the column
density of \ion{O}{1}, damping wings appear before our model
reproduces both sides of the observed spectrum.  To resolve this
discrepancy, we refit the \ion{H}{1}~21-cm absorption lines, using the
Voigt profile fitting code ({\sc MINFIT}; \citealt{chu03}) (instead of
using Gaussian fits as in \citet{kob99}).

These measurements of the \ion{H}{1} column densities require some
assumptions. We calculated them by
\begin{equation}
  N_{HI} = 1.823\times10^{18}\left(\frac{T_S}{f}\right)\int\tau_{21}dV,
\end{equation}
where $T_S$ is the spin temperature in degrees Kelvin, and $\tau_{21}$
is the \ion{H}{1}~21-cm optical depth. We adopted the spin
temperatures from \citet{kob99}. Then by integrating the optical depth
of the model profile that we best fit to the observed spectrum, we
calculated the \ion{H}{1} column densities. Our best fit parameters
are listed in the first three rows of Table~3.  The column densities
are large enough to be classified as sub-DLAs (i.e.,
\logn~(\ion{H}{1}/[\cmm]) $>$ 19)).  The total \ion{H}{1} column
density after adding up the three components will be \nhi\ =
1.27$\times10^{20}$ [\cmm], which is consistent with the values
measured in past studies using different radio telescopes with
different beams on the sky: 1--5$\times10^{20}$ [\cmm] \citep{mat84},
1--2$\times10^{20}$ [\cmm] \citep{kob99}, 1.7$\times10^{20}$ [\cmm]
\citep{kal05}, 1.3$\times10^{20}$ [\cmm] \citep{leh08}, and
7.1$\times10^{20}$ [\cmm] as an upper limit including contributions
from our Galaxy \citep{gia05}. These \ion{H}{1} line widths are
slightly broader than those found previously by \citet{kob99}.  Using
three components with these larger Doppler parameters to fit the
\ion{O}{1}, we found it possible to reproduce the high-velocity side
of \ion{O}{1}~$\lambda$1302 by adjusting its column density to
\logn~(\ion{O}{1}/[\cmm]) $\sim$16, as shown in Figure~4\footnote{We
also cannot reject the alternative solution that instead of a broader
\ion{H}{1} component on the high-velocity side of the system, there
are two components blended together at that velocity.  Such a model,
however, would involve three additional free parameters (i.e.,
position, column density, and line width) for which we do not have
constraints from our radio spectrum.  We adopt the one component model
for simplicity.}.

The only remaining disagreement is an underproduction by the model of
\ion{O}{1}~$\lambda$1302 at the low-velocity side of the absorption
profile.  Because a similar discrepancy is seen in other metal
absorption lines, such as \ion{Mg}{2}, \ion{Fe}{2}, and \ion{Si}{2},
there is probably an additional component at the low-velocity side of
this system. Assuming a fiducial line width of $b$ = 10~\kms\ (i.e.,
the average Doppler parameter of the other three \ion{H}{1}~21-cm
lines is 9.0~\kms), we adjust both the line position and column
density of \ion{O}{1} and find the best combination: $V_{\odot}$ =
160.8~\kms\ and \logn~(\ion{O}{1}/[\cmm]) $\sim$15.

In fact, another strong constraint for this additional component comes
from the radio observation. We can place an upper limit on the
\ion{H}{1}~21-cm line optical depth of this additional component at
the line center, because we did not detect it at the appropriate
wavelength in the \ion{H}{1}~21-cm absorption spectrum.  The S/N-ratio
of the spectrum is $\sim$40 at that wavelength, which corresponds to a
1$\sigma$ optical depth limit of $\tau_{max}$ = $-\ln (1-\sigma)$
$\sim$0.025.\footnote{i.e., a typical uncertainty of our estimates of
\ion{H}{1} column densities is also about 2.5\%.} As described in the
next section, our best photoionization model for this component gives
\logn~(\ion{H}{1}/[\cmm]) of 19.31 and $b$ of 13.0~\kms\ for the
\ion{H}{1} line, whose optical depth at line center would be $\tau$ =
0.022, smaller than the 1$\sigma$ detection limit in the radio
spectrum (Figure~5).  We assume the same spin temperature for this
component as was measured for the closest component at $V_{\odot}$ =
175.8~\kms\ ($T_S$ = 22~$K$). The line parameters for this additional
component, determined from a fit to the \ion{O}{1} are listed in the
fourth line of Table~3. However, we will not consider this component
in the discussion (\S~6) because the fitted line parameters are all
based on an arbitrary Doppler parameter.

\section{Photoionization Modeling}

\subsection{Modeling Procedure}

We briefly summarize our modeling procedure, although it is similar to
previous studies (e.g., Churchill \& Charlton 1999). Using the
photoionization code Cloudy, version 07.02.00 \citep{fer98}, for each
of the five clouds (the fifth one is introduced later) we search for
the best combinations of the fit parameters: 1) ionization parameter
($\log U$ = $\log [n_{\gamma}/n_{H}]$, defined as the ratio of
ionizing photons to the number density of hydrogen in the absorbing
gas) and 2) metallicity (\logz) in solar units.\footnote{The default
solar composition we assume is listed in {\it Hazy 1}, a manual of
Cloudy. For example, [Si/H]$_{\odot}$, [N/H]$_{\odot}$, and
[O/H]$_{\odot}$ are $-4.46$, $-4.07$, and $-3.31$, respectively.}  The
MB is likely to be confined and compressed by its interaction with the
Galactic halo, but this does not produce shock ionization
\citep{bla07}. Therefore, we consider only photoionization.  We
optimize on the observed column densities of \ion{H}{1} or other metal
ions (i.e., \ion{O}{1} and \ion{Si}{4}) that are listed in the first
column of Table~3, that is we require the Cloudy models to produce
those column densities.  For example, in the case of the component at
$V_{\odot}$ $\sim$176~\kms, we always fix the \ion{H}{1} column
density to \logn~(\ion{H}{1}/[\cmm]) = 19.59, and search for the best
values of $\log U$ and \logz in a grid with steps of 0.1~dex, to
reproduce as many absorption lines from other transitions as
possible. We repeat this procedure for the other four components,
individually.  We assume that the absorbers are plane-parallel
structures of constant density, and that they are in photoionization
equilibrium. We initially adopt a solar abundance pattern, but also
explore some variations based on the observational constraints. For
the incident radiation field, we first consider a pure extragalactic
background radiation with contributions from quasars and star forming
galaxies (with a photon escape fraction of 0.1), following
\citet{haa96,haa01}. We also explore the effects of other incident
radiation fields, including fluxes from our Galaxy and the LMC, as
described in the next section.

For a given $\log U$ and \logz, Cloudy computes the column density of
each element in various ionization stages, and the equilibrium gas
temperature, $T$.  We can calculate the Doppler parameter for each
element using $b$ = $\sqrt{b_{T}^2 + b_{turb}^2}$, where $b_{T}$ is
the thermal broadening defined as $b_{T}$ = $\sqrt{2kT/m}$ and
$b_{turb}$ is the broadening from gas turbulence and bulk motion.  We
use the measured $b$ for the transition on which we optimized (listed
in Table~3) in order to calculate $b_{turb}$, which is then applied
for other elements.  Using these derived line parameters from the
model, we synthesize a spectrum, after convolving the instrumental
line spread function of {\it HST}/STIS, and compare it to the observed
spectrum.  The synthesized spectra are compared to the observed
spectrum by eye, because models with metallicity and ionization
parameters that differ only slightly from the best values (e.g. 0.1 or
0.2 dex) would deviate significantly from the observed spectrum (see
Figure~4 of \citealt{mis08}).  Moreover, a formal procedure like
$\chi^2$ fitting is not applicable because the system in the MB has
various (more than 10) transitions that we must consider
simultaneously, and it is very difficult to determine how they should
be weighted in a $\chi^2$ calculation \citep[e.g.,][]{mis08}.

The transitions we optimize are \ion{H}{1} and \ion{O}{1}, and both
have very low ionization potentials (IP = 13.6~eV). As frequently
reported for photoionization models of \ion{Mg}{2} absorbers, the
low-ionization phase clouds that produce \ion{Mg}{2} absorption lines
also produce other low-ionization transitions (e.g., \ion{C}{2},
\ion{Si}{2}, and \ion{Fe}{2}), but not high-ionization transitions
(e.g., \ion{Si}{4} and \ion{C}{4}). An additional high-ionization
phase is almost always required to reproduce these
transitions. Therefore, we repeat line fitting and photoionization
modeling for the high-ionization phase by optimizing high-ionization
transitions (\ion{Si}{4} in this study, as described below). Finally,
we synthesize a model spectrum including both low- and high-ionization
phases, and compare it to the observed spectrum.

As described above, our method for deriving constraints on
metallicities and ionization parameters of the absorbing gas relies on
Voigt profile fitting of separate components of the absorption profile
of a particular transition which we feel is best constrained.  We then
use photoionization modeling to infer the column densities and Doppler
parameters of other transitions in these Voigt profile component
clouds.  Synthesized profiles from these models are compared to the
data in order that we can place constraints on the parameters.
Although this method does rely on the assumptions behind the Cloudy
models, it has some advantages over simply taking the ratios of
apparent column densities of selected transitions in order to
determine metallicity.  First of all, we can separately examine
properties of individual clouds along the line of sight at different
velocities.  We are not averaging these components together, which the
apparent column density method requires since it does not distinguish,
for example, how much of the hydrogen is associated with the separate
clouds.  Secondly, we can use the components determined from
unsaturated lines in order to constrain the properties of those clouds
using other saturated or partially saturated components as well.  The
measured Doppler parameters for the unsaturated lines can be used
along with the temperature given by a given Cloudy model in order to
determine model Doppler parameters of other lines that may be
saturated or blended.  Comparing the model to the shapes of the
observed profiles of these lines, particularly the shapes of the sides
of the profiles, often yields meaningful constraints on parameters.
Models also take into account the appropriate ionization corrections
in each case.  In this way we can determine the range of acceptable
parameters, considering all observational constraints.  Finally, we
are able to consider separate phases of gas, having different
densities and velocities along the line of sight, by comparing model
predictions to the observations, component by component.

\subsection{Alternate Incident Radiation Fields}

Because the Milky Way (MW) and the LMC are located within several tens
of kilo-parsecs of the MB, they also contribute as additional ionizing
radiation sources. Therefore, we consider three alternative incident
radiation fields, (i) extra-galactic background (EGB;
\citealt{haa96,haa01}) plus radiation from the MW at a distance of $D$
= 50~kpc from the MW \citep{fox05a}, (ii) a maximum flux model, with
the EGB, the MW radiation, plus radiation from the LMC with a 30~\%
escape fraction, and (iii) an intermediate case, with EGB, the MW
radiation, plus radiation from the LMC with a 15~\% escape fraction.
\footnote{We do not consider the contribution from local O/B-type
stars in the MB, because it is less likely that our sightline to the
background quasar goes through stellar associations, at least compared
to the sightlines toward stars in the MB. It is also suggested that
O/B-type stars tend to localize in the wing of the SMC
\citep{irw90,bat92}.}  We construct these radiation fields following
the procedure of \citet{bla99,bla02}. Figure~6 shows the strength of
the ionizing radiation from the MW and the LMC as a function of a
distance from the center of the MW. At $D$ $\sim$50~kpc, at which the
MB is located, the contribution from the LMC to the radiation field is
significant. On the other hand, the contribution from the SMC is
negligible because of its low luminosity.

The spectral shapes for each of the four incident radiation field
models on the MB gas are plotted in Figure~7. The radiation from the
MW and the LMC start to dominate the EBR at $\log$($\nu$/[Hz]) $<$
16.1, and strongly dominate over the EBR at $\log$($\nu$/[Hz]) $<$
15.5. Normalizations for these fields, in units of the density of
ionizing photons, and the contribution from the local flux sources (MW
and LMC) compared to the EBR are summarized in Table~4.

\section{Results}

Transitions that are detected in the {\it HST}/STIS spectrum and those
that provide useful limits are shown in Figure~8, from the lowest
(\ion{Mg}{1}, IP = 7.6~eV) to highest (\ion{N}{5}, IP = 97.9~eV)
ionization potentials, following the \lya\ profile in the first
panel. We also display the observed \ion{H}{1}~21-cm absorption
profile \citep{kob99}.  In Figure~8, 0~\kms\ denotes the apparent
optical depth-weighted median of the MB absorption system,
corresponding to a heliocentric velocity of $V_{\odot}$ =
209.9~\kms. A number of absorption lines from various ions (i.e.,
\ion{Mg}{1}, \ion{C}{1}, \ion{O}{1}, \ion{N}{1}, \ion{Mg}{2},
\ion{Mn}{2}, \ion{Fe}{2}, \ion{Si}{2}, \ion{Ni}{2}, \ion{S}{2},
\ion{Si}{4}, and \ion{C}{4}) are detected.  For \ion{N}{5}, we can
place only an upper limit on the equivalent width.  The
\ion{Mg}{2}~$\lambda$2803 profile suffers from a data defect on its
blue side, so we use the \ion{Mg}{2}~$\lambda$2796 profile as the main
\ion{Mg}{2} constraint.  Since the \ion{Mg}{1}~$\lambda$2853 profile
is noisy (i.e. S/N $\sim$3.7 per pixel), it is only used as a loose
constraint on models.

As a starting point for photoionization modeling, we roughly estimate
possible ranges of gas temperature and ionization parameter ($\log
U$).  The observed line widths of all transitions are similar, which
suggests that bulk motion (gas turbulence) is a dominant source of
line broadening. The relatively narrow 21-~cm lines, especially the
one with $b =6.4$~\kms, then imply that the gas temperature is very
low.  Such a low temperature implies that the ionization parameter is
also very small ($\log U$ $\leq$ $-5.0$).  For the metallicity, the
value in the SMC has been estimated as $Z$ = 0.1 -- 0.2 $Z_{\odot}$
\citep{pag78,wel01}.  Therefore, in the following we explore
metallicities and ionization parameters, in steps of 0.1 dex, in the
ranges $\log U$ = $-$7.0 -- $-$4.0 and \logz\ = $-$1.0 -- 0.0 unless
other parameter ranges are suggested.

\subsection{Three \ion{H}{1}~21-cm Clouds}
At first we seek the best photoionization model parameters for the
three \ion{H}{1}~21-cm clouds at \delv\ = $-$34, $-$3, and 26 \kms\
from the system center (i.e., $V_{\odot}$ = 176, 207, and 236~\kms)
that we determined in \S~3, by optimizing a fit to the
\ion{H}{1}~21-cm absorption line.  We require our model to reproduce
the column density of \ion{H}{1} from that fit.  Because the cloud at
\delv\ = $-$3~\kms\ has clear detections in \ion{C}{1}, \ion{N}{1},
\ion{Ni}{2}, and \ion{S}{2}, without blending with other absorption
features, we begin with this cloud.  We can constrain the metallicity
to be $-0.9$ $\leq$ \logz\ $\leq$ $-0.5$/$-0.5$ $\leq$ \logz\ $\leq$
$-0.2$ to avoid over/under--production of \ion{Ni}{2} and \ion{S}{2}.
Therefore, the acceptable metallicity is \logz\ $\sim$$-0.5$.  With
this metallicity, the ionization parameter must be $-6.0$ $<$ $\log U$
$<$ $-4.8$ to avoid over/under production of \ion{C}{1}.  Even our
favored model, with \logz\ $\sim$$-0.5$ and $\log U$ $\sim$$-5.8$, for
this \delv\ = $-$3~\kms\ cloud substantially overproduces the
\ion{N}{1}~$\lambda$1201 absorption. To reconcile the model with the
observed \ion{N}{1}~$\lambda$1201 absorption, we decrease the nitrogen
abundance by a factor of $0.7$ dex compared to its solar abundance
pattern. The production mechanisms of nitrogen are poorly understood.
However, a nitrogen deficiency in the SMC has already been reported
(e.g., \citealt{mal01}), and similar deficiencies are seen in some DLA
systems (e.g., \citealt{pet02}) and weak \ion{Mg}{2} systems
\citep{zon04}.  Since \ion{O}{1} and \ion{H}{1} are strongly coupled
in terms of their ionization, their ratio is often used as a measure
of metallicity \citep{leh08}.  In our case, the \ion{O}{1} is highly
saturated so that the Voigt Profile components cannot be measured
directly from the profile.  However, once we have determined the
metallicity from the weaker \ion{Ni}{2} and \ion{S}{2} profiles, we
can verify that the favored model produces an \ion{O}{1}/\ion{H}{1}
ratio consistent with these metallicities.  For the $-$3~\kms\
component we find $\log$[N(\ion{O}{1})/N(\ion{H}{1})] = $-$3.81,
corresponding to \logz = $-$0.5. The fact that this is consistent with
the metallicity inferred from the weaker profiles confirms that
ionization corrections are inferred properly from our models, and that
the saturated \ion{O}{1} profile has been separated self-consistently
into Voigt profile components.

Next, we consider the \ion{H}{1} cloud at \delv\ = 26~\kms.  Because
this component has the largest recessional velocity, it should
reproduce at the high-velocity side of all absorption features seen in
the \ion{O}{1}, \ion{Mg}{2}, \ion{Fe}{2}, and \ion{Si}{2} lines.  We
also require that the component reproduces the weak \ion{Ni}{2}
absorption.  The \ion{Si}{2} provides the best lower limit on
metallicity, \logz\ $>$ $-0.8$ because the less saturated
\ion{Si}{2}~$\lambda$ 1304 profile is available.  Similarly, an upper
limit of \logz $<$ $-0.6$ applies in order that \ion{Ni}{2} is not
over-produced.  At the preferred value of \logz\ $\sim$$-0.7$, there
is a strict lower limit on the ionization parameter of $\log U$ $>$
$-6.1$ under which \ion{C}{1} would be over-produced. We do not place
any formal upper limit on the ionization parameter of this cloud,
since the constraints from the relatively low signal-to-noise spectrum
were not significant. However, we can place a marginal upper limit,
$\log U$ $<$ $-5.0$ to avoid an under-production of \ion{Si}{2}, once
we adopt \logz\ = $-0.7$. We favor values toward the lower end of this
range, $\log U$ $\sim$$-6.0$, because the fits to \ion{O}{1},
\ion{Mg}{2}, and \ion{Fe}{2} are slightly better. We also find that
this cloud must have a deficiency of nitrogen of $1.4$ dex compared to
the solar value, although this value could be smaller (\S.~6).

Because the third \ion{H}{1} cloud at \delv\ = $-$34~\kms\ is heavily
blended with the adjacent clouds, it is difficult to place strict
constraints on its physical conditions.  Therefore, we simply note
that its metallicity and ionization parameter could be very close to
those of the cloud at \delv\ = 26~\kms. Several transitions,
\ion{Mg}{1}, \ion{Ni}{2}, and \ion{C}{1}, are over-produced for
metallicities higher than \logz\ $>$ $-0.6$ or $-0.7$.  The
metallicity should also be greater than $-1.0$, to avoid an
under-production of \ion{Si}{2}. Although we cannot place strong
constraints on the ionization parameter, the $\log U$ should be
between $-6.0$ and $-5.0$ in the acceptable range of the metallicity,
$-1.0$ $<$ \logz\ $<$ $-0.7$, to reproduce the observed profiles of
\ion{C}{1}, \ion{Si}{2}, and \ion{O}{1}. We adopt a model with $\log
U$ $\sim$$-5.7$ and \logz\ $\sim$$-0.7$, the same as for the \delv\ =
26~\kms\ cloud, because it is consistent with the observations.  We
also find that this cloud must have a deficiency of nitrogen of $1.0$
dex compared to the solar value.

\subsection{High Ionization Cloud}
In addition to the three \ion{H}{1} clouds above, one high-ionization
cloud, at \delv\ = 7~\kms\ (i.e., $V_{\odot}$ = 217~\kms), is
necessary to reproduce absorption in the high-ionization lines,
\ion{Si}{4} and \ion{C}{4}.  To determine the photoionization model
parameters for this cloud, we optimizes on the \ion{Si}{4} column
density, since the Voigt profile fitting is better in this region than
it is for the \ion{C}{4} doublet.  The result of a Voigt profile fit
to the \ion{Si}{4} is given in the last row of Table~3. Comparing to
the observed \ion{C}{4} absorption, we find an ionization parameter of
$-2.7$ $\leq$ $\log U$ $\leq$ $-2.4$.  The metallicity of the
high-ionization cloud is constrained to be \logz\ $>$ $-4.0$ by the
requirement that the corresponding \ion{H}{1}~21-cm line is not
detected at that velocity.  The metallicity could actually be much
higher, as high as that of the low ionization clouds, but we have no
way to place further constraints.

\subsection{Additional \ion{O}{1} Cloud}
A model with three low ionization clouds and one high-ionization cloud
reproduces the observed spectrum very well except for the low-velocity
side of \ion{O}{1}, \ion{Fe}{2}, and \ion{Si}{2}.  As noted in \S~3,
we add an additional cloud at \delv\ = $-$49~\kms\ (i.e., $V_{\odot}$
= 161~\kms) in order to fully produce the observed \ion{O}{1}
absorption.  For this cloud, we can constrain the metallicity by the
requirement that the optical depth of \ion{H}{1}~21-cm at line center
must not exceed $\tau$ $=$ $0.025$.  We obtain \logz\ $\geq$ $-1.0$
for $\log U$ $\geq$ $-5.1$, and \logz\ $\geq$ $-0.9$ for $\log U$
$\leq$ $-5.2$.  The ionization parameter should be $\log U$ $>$
$-6.0$, below which \ion{Mg}{1} would be overproduced. An upper limit
on the ionization parameter of this component is $-3.5$, to avoid an
over-production of \ion{Si}{4}.  We again see that nitrogen is
deficient in this cloud by $0.6$ dex.  We give a sample of an
acceptable model for this cloud, with \logz\ = $-1.0$ and $\log U$ =
$-5.1$ in Table~3.  However, we note that the parameters could instead
be more similar to those of the other clouds.

\subsection{Our Best Model}
Table~3 lists the best parameters of our photoionization models:
column (1) is the optimized transition, column (2) is the heliocentric
velocity of the absorption line, column (3) is relative velocity from
the system center, columns (4) and (5) are absorption optical depth at
the line center and spin temperature measured from the
\ion{H}{1}~21-cm emission line (only for \ion{H}{1}~21-cm lines),
columns (6) and (7) are the best Voigt profile fit values of column
density and Doppler parameter, columns (8) and (9) are the best model
parameters (assuming a solar abundance pattern) of metallicity and
ionization parameter, column (10) is gas temperature from our model,
columns (11) and (12) are the gas volume density and the thickness of
the absorber assuming a plane-parallel structure, and column (13)
notes changes in the nitrogen abundance (compared to the Solar
abundance pattern) that are necessary to reproduce the observed
spectrum. The synthesized spectrum, using the best model parameters,
is overplotted on the observed spectrum for each transition in
Figure~8. If we compare the models with/without nitrogen deficiency to
the observed spectrum, it is obvious that the nitrogen deficiency is
necessary, as shown in Figure~9.  A summary of how specific
transitions were used to constrain $\log U$ and \logz\ for the MB
system is given in Table~5. We also compare the equivalent widths of
the observed and the modeled spectra in Table~2, and confirmed they
are in good agreement.

\subsection{Effects from Additional Radiation Sources}
Because this absorber is located in the MB, close to the MW and the
LMC, the radiation field around it could be enhanced by contributions
from these galaxies, as mentioned in \S~4.2. We perform
photoionization analysis using the various radiation fields listed in
Table~4. Because the purpose of this analysis is to examine how the
shape of radiation field affects the results of the photoionization
modeling, we always use the best-fit parameters from Table~3.  We
compare the model spectra using various incident radiations to the
observed spectrum in Figure~10. On the observed spectrum, we do not see
any significant differences between the results for the different
radiation field models, except for \ion{C}{1}~$\lambda$1329, for which
the model is slightly improved if we take a radiation from the LMC
into account. However, the quality of spectrum around this line is not
very high. Apparently the same fit parameters are still acceptable for
the alternative radiation fields. However, once we estimate the line
parameters of various transitions in the strongest component at \delv\
= $-$3~\kms\ when using these incident fields, there are noticeable
differences for some transitions (i.e., \ion{C}{4}, \ion{N}{3},
\ion{Al}{3}, and \ion{Si}{4}) that are very weak and/or positioned at
low-S/N regions. This is because the number of hydrogen ionizing
photon is basically same by our assumption (because we assume the same
ionization parameter), however, the ionizing photons of some ions
(especially ions with low ionization potentials) would be increased
significantly once the additional radiation is considered.  Although
the ionization parameter that we infer is the same for these
alternative radiation fields, the number of ionizing photons is
substantially increased so that this same ionization parameter
corresponds to a higher density.  For the \delv\ = $-$3~\kms\
component the density has increased from 0.56~\cmmm\ to 19~\cmmm, and
the size decreased from 25~pc to 0.72~pc in the most extreme case.
The other clouds are similarly smaller, with the \ion{Si}{4} cloud
reduced in size to 0.57~kpc under the more intense LMC radiation
field.  Since the differences in the ionization parameters that we
infer are so small between the different models, it is not practical
to distinguish between the different possible radiation fields using
these data, and thus the inferred densities and sizes are uncertain.
Hereafter, we discuss the results using only the EGB radiation, as
summarized in Table~3.

\section{Discussion}

\subsection{Origin of Metal-enrichment in the MB}
There have been three scenarios prosed to explain the current chemical
composition of the MB.  Tidal stripping from the SMC, purported to
explain the origin of the MB \citep{gar96} would imply a chemical
composition similar to the SMC itself, or perhaps slightly more metal
poor if the MB is preferentially formed from material drawn from the
comparatively pristine outskirts of the SMC. The presence of any
chemical gradient in the SMC is not yet well defined, but the recent
results of \citet{car08} suggest it may be non-negligible.  A galactic
wind from the SMC could enrich the MB with metals, and such a wind
would likely be enhanced in alpha elements, given the composition of
supernova-driven winds in other dwarf galaxies where such signatures
are observed several kpc from their origin \citep{mar02, str04}.  The
final possibility is in-situ enrichment from stars formed in the MB
\citep{dem98}.  Given the $\sim$200~Myr age of the MB \citep{gar96},
there has been ample time for several generations of the most massive
B stars to evolve and contribute their nucleosynthetic products to
their surroundings.  Some combination of these processes are likely to
play a role, but the balance of these has not yet been determined.

\citet{rol99} found an underabundance ($\sim$0.5~dex) in the light
metals of a B-type star in the MB, DGIK~975, compared to the SMC
\ion{H}{2} regions. Similarly, metallicity of the interstellar medium
in the MB toward a young star, DI~1388, was measured to be
$\sim$0.2~dex lower than the SMC \ion{H}{2} regions \citep{leh08}.
The sightline toward our target, PKS~0312-770, has an angular
separation of $\sim$4.1$^\circ$ from DI~1388, which corresponds to
about $\sim$4.4~kpc in physical scale at the distance of the SMC ($d$
$\sim$60~kpc).  However, we have found that the metallicity in the MB
toward PKS~0312-770 is higher than that measured toward DI~1388 by
$\sim$0.5~dex, and even $\sim$0.1 -- 0.2~dex more metal-rich than the
SMC \ion{H}{2} regions. What is the source of this difference?  One
possible idea is that the sightline toward PKS~0312-770 is not mixed
with metal-poor gas, like the sightline toward DI~1388. If this is the
case, absorbers with lower-metallicities should have higher total
hydrogen column densities. Such a trend (i.e., a gradual increase in
metallicity with decreasing hydrogen column density) has already been
pointed out for various objects at higher redshift, including
galaxies, (sub-)DLA systems, and the intergalactic medium (e.g.,
\citealt{boi98,per03,yor06,mis08}).

There are observational results that support the scenario above.
Column densities of sulfur (a modestly dust-depleted element) are
similar between these two sightlines.  Each of the three clouds toward
PKS~0312-770 (with \logn~(S/[\cmm])is 14.2, 14.4, and
14.2)\footnote{We fitted the \ion{H}{1}~21-cm absorption profile with
three components. The total \ion{S}{2} column density
($\log$(\ion{S}{2}) $\sim$14.75) is slightly lower than that measured
in \citet{leh08} because we assume \ion{S}{2}~$\lambda$1254 is blended
with an unrelated line to avoid over-absorption of
\ion{S}{2}~$\lambda$1251, for which \citet{leh08} applied the AOD
method directly.} has a similar column density to the total
(\logn~(S/[\cmm]) = 14.35) in DI~1388. However, the PKS~0312-770
clouds have total hydrogen column densities (\logn~(H/[\cmm]) = 19.6,
19.6, 19.7)\footnote{Note that neutral hydrogen column densities
toward DI~1388 and each of the three clouds toward PKS~0312$-$770 are
almost the same (\lognhi\ $\sim$19.6), but the former is considerably
more ionized (\lognhii\ $\sim$20.0) than the latter (\lognhii\
$\sim$17.7).} 0.6 dex smaller than the total in DI~1388
(\logn~(H/[\cmm]) = 20.2).  
This implies a higher metallicity and less dilution in PKS~0312-770.
These results are consistent with the scenario above: i.e.,
metal-enriched gas flowed out toward the MB from the SMC almost
isotropically, but the MB absorbers toward PKS~0312$-$770 were only
weakly diluted by metal-free material, while the absorber toward
DI~1388 is significantly diluted.

\subsection{Nitrogen Deficiency in the MB}
The overall metallicity and the chemical abundance pattern in the MB
can be used to infer the origin of this inter-Cloud
material. \citet{ham94} and \citet{rol99} measured He, O, Si, Mg, and
N abundances for several early B stars in the MB, assuming that their
composition reflects the present-day ISM, and concluded that these
stars were 0.5 dex more metal-poor than SMC stars. These authors
suggested that the MB stars formed from a mixture of SMC and
un-enriched gas.  \citet{duf84} analyzed the \ion{H}{2} region He, C,
N, O, and Si abundances for the SMC, finding that the overall
metallicity, as measured by the alpha elements O and Si, is $\sim$0.5
dex more metal poor than the Sun.  The N/Si ratio in \ion{H}{2}
regions is 0.5 dex lower than the Sun, consistent with the trend
observed in other low-metallicity dwarf galaxies (e.g.,
\citealt{ks96,nav06}). We summarize these measurements from the
literature along with ours in Table~6. Locations of the MB stars are
also shown in Figure~1.

Figure~11 shows the abundance ratio $\log$(N/Si) versus $\log$(Si/H)
for the aforementioned measurements from Table~6, assuming a solar
abundance pattern if another element (instead of silicon) is used to
estimate these parameters. The figure includes our three velocity
components toward PKS~0312-770 (open circles), the \citet{rol99} MB
stars (filled stars), the gas toward the MB star DI~1388
(\citet{leh08}; open star), the SMC star AV~304 (open square), the Sun
(solar symbol), the range of SMC \ion{H}{2} regions (filled square),
the dwarf galaxies from \citet{nav06} (crosses; values computed from O
abundance measurements assuming a solar Si/O ratio), and damped Lyman
alpha systems from \citet{hen07} (dots).  The abscissa reflects the
overall alpha element abundance of the systems in question and shows
that the metallicities of the SMC star AV~304, the MB stars, and the
MB gas toward PKS~0312-770 are generally consistent with the range of
metallicities in SMC \ion{H}{2} regions.  The MB star DGIK~975 is a
possible exception, appearing $\sim$0.5 dex more metal poor than the
rest of these measurements.  Figure~3 of \citet{rol99} shows that this
star also lies the furthest from the SMC, in a low-\ion{H}{1} column
density region equidistant between the SMC and LMC.

The N/Si ratio shown on the ordinate of Figure~11 is a measure of the
chemical enrichment timescale.  Si is synthesized in massive stars and
returned to the interstellar medium through supernovae on timescales
of $\sim$10 Myr.\footnote{However, there is the possibility that
$\alpha$-elements released by supernovae remain for an extended period
in hot $10^6$~K bubbles and require several hundred Myr to mix with
ambient galactic material \citep{ten96,ks97}.} Nitrogen, by contrast,
is thought to be produced in most galaxies by low- and
intermediate-mass stars and released on timescales of $>$100~Myr
\citep{vdh,marigo,rv81}.\footnote{See, however,
\citet{ack04,spi05,pet08} for evidence that massive stars may dominate
N production at {\it very} low metallicities, $Z$ $<$
0.01~$Z_{\odot}$.}  Thus, the N/Si ratio may drop during prolonged
starbursts and rise during prolonged periods of quiescence,
functioning as a kind of ``clock'', marking the time since the most
recent major episode of star formation \citep{ep78,pan88,gar90,ks98}.
\citet{hen07} used this scenario to model the chemical evolution of
DLA systems in the N/Si vs. Si/H plane and found that they could
reproduce the properties of most DLA systems as a function of two
variables: the star formation efficiency and the age of the system.
They concluded that small ages (i.e., less than 250~Myr) are required
to produce the low N/Si systems, while high star formation
efficiencies lead to higher Si/H and lower N/Si ratios. However, a
number of other processes may contribute to shaping the chemical
evolution of a system.  The arrows in Figure~11 show qualitatively the
four possible evolutionary vectors attributed to N enrichment
(vertically upward), $\alpha$ enrichment (toward the lower right),
dust depletion (toward upper left), and dilution by metal-poor gas
(leftwards).

Figure~11 shows that the 207~\kms\ component toward PKS~0312-770 is
0.3--0.4 dex more metal-rich than the bulk of the SMC and other MB
stars. This component also exhibits an N/Si ratio that is 0.2 -- 0.3
dex lower than the SMC stars and \ion{H}{2} regions.  This departure
is consistent with a composition consisting of SMC material augmented
by a small amount of $\alpha$ enrichment from supernovae.  None of the
MB stars in Figure~11 nor the ISM probed toward the MB star DI~1388
share this abundance pattern.
\footnote{Hot stars in the MB could be affected by additional
production of nitrogen.}  The PKS~0312-770 line of sight is therefore
the first MB location with a metallicity that is similar to, or
perhaps slightly higher than, the SMC at large.

Figure~11 shows that the absorbing component at 236~\kms\ has a
metallicity similar to the SMC and most MB stars, but the N/Si ratio
is about 1.0 dex lower.\footnote{Normalization of the spectrum may be
performed incorrectly to the low-S/N ($\sim$5) spectrum region around
\ion{N}{1} lines. If we reduce nitrogen abundance by only $-1.0$ dex
from the solar abundance pattern (i.e., same as the 176~\kms\
component), \ion{N}{1}~$\lambda$1200b and \ion{N}{1}~$\lambda$1201
would be reproduced better (see Figure~12).} Such a low N/Si ratio
places this material among the most N-deficient damped Lyman alpha
systems and suggests a nucleosynthetic history dominated by
$\alpha$-producing massive stars.  In the \citet{hen07} models, such a
low N abundance at such high metallicity can only be achieved by very
efficient and very recent star formation such that massive stars
dominate the mass-averaged nucleosynthetic contribution with virtually
no contribution from longer-lived N-producing stars.  Given the
presence of B stars in the MB (e.g., \citealt[][]{rol99}), some
supernova activity during the $<$ 200~Myr lifetime of the MB is
likely.  In situ enrichment appears a plausible explanation for this
component on the basis of chemical cloud.
However, it is also possible that a Si enrichment of ambient SMC
material plus dilution from putative metal-poor halo gas could also
explain this data point (i.e., a combination of vectors that together
drive evolution in Figure~11 in a downward vertical direction from the
SMC \ion{H}{2} regions composition).

Finally, the 176~\kms\ component has a Si/H ratio consistent with the
SMC and the 236~\kms\ component, but the uncertainties on the N/Si
ratio are too large to place meaningful constraints on the origin of
this material.  Moreover, abundance pattern would also be strongly
affected by dust depletion and ionization conditions, which can only
be explored with very high-S/N spectrum.  Therefore, this component
may either be N-deficient or it may be consistent with the SMC
\ion{H}{2} regions.

\subsection{Dust Depletion in the MB}

In addition to various absorption lines we have detected in the {\it
HST}/STIS spectrum, \citet{smo05} also detected the \ion{Ca}{2}~K line
in their medium-resolution ($R$ = 6000) optical spectra of 7 quasars
behind the MB including our target \citep{smo05}.  By comparing the
total column densities of \ion{Ca}{2} and \ion{H}{1} (measured from
\ion{H}{1}~21-cm emission line), they found that the abundance ratio
of \ion{Ca}{2} to \ion{H}{1} in the MB is systematically higher than
that of Galactic gas by a factor of $\sim$0.5 dex. \citet{smo05}
proposed possible scenarios for this difference, such as the higher
ionization condition of the hydrogen gas, and weaker dust-depletion in
the MB.  To test the scenario, a higher-resolution spectrum will be
necessary to deblend an unresolved \ion{Ca}{2}~K profile into multiple
components (as we did for other transitions in the UV spectrum) in
order to constrain the photoionization model.

In \S~5, we found the best model parameters. However, there is still
an ambiguity due to the possible effects of dust, which would lead to
different inferred physical conditions.  Our data was not of
sufficient quality to measure dust depletion, but the estimated low
gas temperature of the three \ion{H}{1} clouds ($T_{gas}$ $<$ 1000~K)
would imply that dust grains could survive in the absorber. Moreover,
the \ion{Ca}{2}~K absorption strength, expected from our best model
($W_{rest}$=0.99\AA), is seven times greater than the observed value
toward PKS~032$-$770 ($W_{rest}$=0.14\AA; \citealt[]{smo05}). This
also implies that the absorber contains substantial amounts of dust
because calcium is one of the most severely depleted elements
\citep{sav96}.  Actually, \citet{leh08} suggested Si and Fe in the MB
toward DI~1388 are depleted to dust by factors of $-0.45$ and $-0.61$
dex respectively, although the condition could be different toward
PKS~0312$-$770.  If depletion onto dust is significant, then the Si
abundances and the implied metallicities become even larger, making
this sightline significantly more metal-enriched than the SMC itself.

\subsection{Comparison with Damped Ly$\alpha$ Systems at high-$z$}

Damped Ly$\alpha$ systems (DLAs) are characterized by high \ion{H}{1}
column densities (i.e., \logn~(\ion{H}{1}/[\cmm]) $\geq$ $2 \times
10^{20}$~\cmmm) and low metallicities (i.e., $Z$ $\sim$0.1 -- 0.01
$Z_{\odot}$) (e.g. \citealt[][]{pet97,pro03}).  DLAs are also known to
have lower nitrogen abundance relative to $\alpha$-elements, compared
to solar abundance pattern (e.g., \citealt[][]{pet02}). DLAs provide
plentiful information on the physical condition in gas clouds in the
ancient universe that can never be traced by stellar objects.
However, so far DLA absorbers have not been identified clearly,
especially at higher redshift, although at least elliptical galaxies
are probably ruled out \citep{cal03}. Thus, it is quite helpful to
find local counterparts of those systems and study their properties in
detail.

With respect to \ion{H}{1} column density, the MB absorbers toward
PKS~0312$-$770, whose total column density is just below the criterion
to be classified as a DLA system, could be analogs of high-$z$ DLA
systems. However, their nitrogen abundance relative to
$\alpha$-elements (e.g., [N/Si]) tend to be small and positioned at
the lowest end of the [N/Si] distribution of DLA systems in
Figure~11. Particularly, the nitrogen abundance of the $V_{\odot}$ =
236 cloud is very small compared to that expected for DLA systems and
blue compact galaxies of a similar metallicity.  The origin of this
difference should be understood before using the MB absorbers as local
counterparts of high-$z$ DLA systems.

The nitrogen deficiency may be linked to the synthesis process. It is
already known that nitrogen is produced by low and intermediate mass
stars (LIMS; $M/M_{\odot}$ $\leq$ 8), while $\alpha$-elements are
produced by massive stars. For DLA systems with low nitrogen
(hereafter, low nitrogen DLAs: LNDLAs), there have been two possible
scenarios presented (\citealt[][]{hen07} and references therein):
i.e., (i) they are in early production stages of nitrogen released
from LIMS (delay scenario, \citealt[][]{pet02}), and (ii) they have an
intrinsically flattened or truncated initial mass function (IMF) with
fewer LIMS (reduction scenario, \citealt[][]{pro02}).  However, as
summarized in \citet{hen07}, neither scenario can explain the nitrogen
deficiency of LNDLAs perfectly; the former cannot reproduce a possible
bimodality of [N/$\alpha$] distribution of DLAs \citep{pro02,cen03},
while the latter produces too much iron compared with the observed
amount \citep{lan03}.

Thus, we have not yet identified the origin of the nitrogen
deficiency.  Nonetheless, it is likely that the production history of
nitrogen is not the same between the SMC and the MB.  The IMF of a
stellar association NGC~602 in the SMC (a single power-law with a
slope of $\Gamma$ $\sim$$-1.2$ for $M/M_{\odot}$ = 1 -- 45;
\citealt[][]{sch08}) is quite similar to the IMF in the solar
neighborhood ($\Gamma$ $\sim$$-1.35$ for $M/M_{\odot}$ = 0.4 -- 10;
\citealt[][]{sal55}). This means that the abundance patterns of the MW
and the SMC are expected to be similar, while they can be different
from that of the MB, whose IMF is not necessarily the same. The MB
probably has a local star forming history independent of the nearby
star forming regions (i.e., the LMC and the SMC), which is consistent
with the observed results that the young stars in the MB are not old
enough to have escaped from the SMC based on their peculiar motions.

If the MB absorbers are indeed local counterparts of high-$z$ DLAs,
our results would have interesting implications: (i) DLA systems along
our sight-lines to the background quasars do not necessarily represent
global properties of absorbing structures, and (ii) the large scatter
of metallicity and [N/$\alpha$] values could be due to the internal
gradient of each DLA absorber, and they are different from place to
place along different sight-lines go through. For example, we would
under-estimate the global metallicity (and nitrogen abundance) of the
Magellanic Clouds if our line of sight went through only the
MB. Moreover, there might be metallicity gradient in the MB itself;
higher in the SMC wing than in the remaining parts of the MB
(\citealt[][]{leh08} and references therein). As for DLAs,
\citet{ell05} and \citet{che05} already suggested that there could be
a metallicity gradient as a function of a distance from the galactic
center. \citet{ell05} and \citet{wol03} also proposed that the DLA
region is sampling a lower metallicity than the star-forming
regions. Thus, high-$z$ DLA systems also may have complex internal
structures like the Magellanic Clouds and their neighborhood.

\acknowledgments This research was funded by the National Science
Foundation (NSF) under grant AST 04-07138 and by NASA under grant
NAG5-6399.  TM acknowledges support from the Special Postdoctoral
Research Program of RIKEN. This work was supported by NASA through
grant NAG5-10770 to HAK.  BPW was supported by HST grant
HST-R-10984.01-A and NASA ADP grant NNX07AH42G.  JBH is supported by a
Federation Fellowship from the Australian Research Council. We would
like to thank Nicolas Lehner and the anonymous referee for their
valuable comments.

\clearpage


\begin{deluxetable}{ccccccccrc}
\tabletypesize{\scriptsize}
\tablecaption{HST/STIS Observation Log \label{t1}}
\tablewidth{0pt}
\tablehead{
\colhead{QSO}           &
\colhead{RA}            &
\colhead{Dec}           &
\colhead{$z_{em}$}      &
\colhead{$m_{V}$}       &
\colhead{Date}          &
\colhead{grating}       &
\colhead{$\lambda$-coverage} &
\colhead{t$_{exp}$}          &
\colhead{HST Dataset ID}     \\
\colhead{}              &
\colhead{(h:m:s)}       &
\colhead{(d:m:s)}       &
\colhead{}              &
\colhead{(mag)}         &
\colhead{(yyyy mm dd)}  &
\colhead{}              &
\colhead{(\AA)}         &
\colhead{(sec)}         &
\colhead{}              \\
\colhead{(1)}           &
\colhead{(2)}           &
\colhead{(3)}           &
\colhead{(4)}           &
\colhead{(5)}           &
\colhead{(6)}           &
\colhead{(7)}           &
\colhead{(8)}           &
\colhead{(9)}           &
\colhead{(10)}          
}
\startdata
PKS~0312$-$770 & 03:11:55.4 & $-$76:51:50.8 & 0.2230 & 16.1 & 2001 05 12 & E140M & 1150 -- 1729 & 12629 & O65T01010 \\
            &            &               &        &      & 2001 03 07 & E140M & 1150 -- 1729 & 12629 & O65T02010 \\
            &            &               &        &      & 2001 10 16 & E140M & 1150 -- 1729 & 12650 & O65T13010 \\
            &            &               &        &      & 2001 10 14 & E230M & 2132 -- 2984 &  6060 & O65T14010
\enddata 
\end{deluxetable}

\begin{deluxetable}{lcc}
\tabletypesize{\scriptsize}
\tablecaption{Observed and Modeled Rest-Frame Equivalent Widths \label{t2}}
\tablewidth{0pt}
\tablehead{
\colhead{Transition}           &
\colhead{$W_{rest}$(spec)$^a$} &
\colhead{$W_{rest}$(mod)$^b$}  \\
\colhead{}                     &
\colhead{(m\AA)$^c$}           &
\colhead{(m\AA)$^c$}           \\
\colhead{(1)}                  &
\colhead{(2)}                  &
\colhead{(3)}                  
}
\startdata
 \lya                       & $<$ 14.73 \AA       $^d$ & 8.17 \AA\\
 \ion{Mg}{1}~$\lambda$2853  &      496 $\pm$   51      &  685 \\
 \ion{C }{1}~$\lambda$1329  &       18 $\pm$    4      &   83 $^e$ \\
 \ion{O }{1}~$\lambda$1302  &      520 $\pm$    7      &  496 \\
 \ion{N }{1}~$\lambda$1200b & $<$  311 $\pm$   12 $^d$ &  278 \\
 \ion{N }{1}~$\lambda$1200a &      201 $\pm$   20      &  235 \\
 \ion{N }{1}~$\lambda$1201  &      169 $\pm$   13      &  179 \\
 \ion{Mg}{2}~$\lambda$2796  &     1197 $\pm$   42      & 1108 \\
 \ion{Mg}{2}~$\lambda$2803  &      904 $\pm$   43      & 1063 \\
 \ion{Mn}{2}~$\lambda$1199  & $<$  311 $\pm$   12 $^d$ &  279 \\
 \ion{Mn}{2}~$\lambda$2594  &            $<$   49 $^f$ &  134 \\
 \ion{Fe}{2}~$\lambda$1261  & $<$  694 $\pm$    7 $^d$ &  534 \\
 \ion{Fe}{2}~$\lambda$1608  &      463 $\pm$   21      &  410 \\
 \ion{Fe}{2}~$\lambda$2344  &      828 $\pm$   29      &  765 \\
 \ion{Fe}{2}~$\lambda$2374  &      539 $\pm$   42      &  546 \\
 \ion{Fe}{2}~$\lambda$2383  &      806 $\pm$   56      &  876 \\
 \ion{Fe}{2}~$\lambda$2587  &      835 $\pm$   44      &  780 \\
 \ion{Fe}{2}~$\lambda$2600  &     1059 $\pm$   58      &  938 \\
 \ion{Si}{2}~$\lambda$1190  &      490 $\pm$   16      &  412 \\
 \ion{Si}{2}~$\lambda$1193  &      495 $\pm$   14      &  444 \\
 \ion{Si}{2}~$\lambda$1260  & $<$  694 $\pm$    7 $^d$ &  533 \\
 \ion{Si}{2}~$\lambda$1304  &      451 $\pm$    8      &  423 \\
 \ion{Si}{2}~$\lambda$1527  &      563 $\pm$    9      &  538 \\
 \ion{Ni}{2}~$\lambda$1317  &       26 $\pm$    4      &   98 \\
 \ion{S }{2}~$\lambda$1251  &       40 $\pm$    8      &   42 \\
 \ion{S }{2}~$\lambda$1254  &      120 $\pm$    7      &   78 \\
 \ion{S }{2}~$\lambda$1260  & $<$  426 $\pm$    7 $^d$ &  110 \\
 \ion{C }{2}~$\lambda$1335  & $<$  841 $\pm$    8 $^d$ &  521 \\
 \ion{Si}{3}~$\lambda$1207  & $<$  999 $\pm$   13 $^d$ &  300 \\
 \ion{Si}{4}~$\lambda$1394  &       78 $\pm$    7      &  101 \\
 \ion{Si}{4}~$\lambda$1403  &       73 $\pm$    8      &   57 \\
 \ion{C }{4}~$\lambda$1548  &       94 $\pm$   12      &   95 \\
 \ion{C }{4}~$\lambda$1551  &       67 $\pm$   14      &   51 \\
 \ion{N }{5}~$\lambda$1239  &            $<$   12 $^f$ &  2.3 \\
 \ion{N }{5}~$\lambda$1243  &            $<$   10 $^f$ &  1.1 \\
 \ion{Ca}{2}~$\lambda$3935  &      142 $\pm$   23 $^g$ &  990 \\
\enddata 
\tablenotetext{a}{rest-frame equivalent width or detection limit.}
\tablenotetext{b}{rest-frame equivalent width of best model (see \S~4).}
\tablenotetext{c}{except for \lya\ for which the unit is Angstroms.}
\tablenotetext{d}{blending with other lines.}
\tablenotetext{e}{this would be 46.8~m\AA\ if we assume most likely
                  radiation field, i.e., including extra radiation
                  from the MW and the LMC with 30\% escape fraction
                  (see \S~5).}
\tablenotetext{f}{not detected with 5$\sigma$.}
\tablenotetext{g}{detected in lower resolution spectrum ($R$ $\sim$
                  6000) taken with NTT/EMMI by \citep{smo05}.}
\end{deluxetable}
\clearpage

\begin{deluxetable}{ccccccccccccccc}
\rotate
\tabletypesize{\scriptsize}
\tablecaption{Best Fit Model for the MB Absorber toward PKS~0312-770\label{t3}}
\tablewidth{0pt}
\tablehead{
\colhead{Transition$^a$}            &
\colhead{$V_{\odot}$$^b$}           &
\colhead{$\Delta v$$^c$}            &
\colhead{$\tau_{cen}$$^d$}          &
\colhead{$T_{S}$$^e$}               &
\colhead{$\log N$$^f$}              &
\colhead{$b$$^g$}                   &
\colhead{$\log$(Z/Z$_{\odot}$)$^h$} &
\colhead{$\log U$$^i$}              &
\colhead{$T_{gas}$$^j$}             &
\colhead{$n_{H}$$^k$}               &
\colhead{size$^l$}                  &
\colhead{N deficiency$^m$}          &
\colhead{$\log$(Si/H)}              &
\colhead{$\log$(N/Si)}              \\
\colhead{}              &
\colhead{(\kms)}        &
\colhead{(\kms)}        &
\colhead{}              &
\colhead{(K)}           &
\colhead{(\cmm)}        &
\colhead{(\kms)}        &
\colhead{}              &
\colhead{}              &
\colhead{(K)}           &
\colhead{(\cmmm)}       &
\colhead{(kpc)}         &
\colhead{}              &
\colhead{}              &
\colhead{}              \\
\colhead{(1)}           &
\colhead{(2)}           &
\colhead{(3)}           &
\colhead{(4)}           &
\colhead{(5)}           &
\colhead{(6)}           &
\colhead{(7)}           &
\colhead{(8)}           &
\colhead{(9)}           &
\colhead{(10)}          &
\colhead{(11)}          &
\colhead{(12)}          &
\colhead{(13)}          &
\colhead{(14)}          &
\colhead{(15)}          \\
}
\startdata
\ion{H}{1}$^n$  & 175.8 & $-$34.1 & 0.087 & 22 & 19.59 &  6.4 & $-$0.7 & $-$5.7 &   583 & 0.44    &  0.029 & $-$1.0 & $-5.2$ & $-0.6$ \\
\ion{H}{1}      & 207.3 &  $-$2.6 & 0.049 & 29 & 19.61 &  8.8 & $-$0.5 & $-$5.8 &   287 & 0.56    &  0.025 & $-$0.7 & $-5.0$ & $-0.3$ \\
\ion{H}{1}      & 236.1 &    26.2 & 0.027 & 46 & 19.68 & 11.9 & $-$0.7 & $-$5.7 &   501 & 0.44    &  0.035 & $-$1.4 & $-5.2$ & $-1.0$ \\
\ion{O}{1}$^o$  & 160.8 & $-$49.1 &       &    &(15.00)&(10.0)& $-$1.0 & $-$5.1 &  4450 & 0.11    &  0.065 & $-$0.6 & $-5.5$ & $-0.2$ \\
\ion{Si}{4}$^p$ & 216.8 &     6.9 &       &    & 13.16 & 22.1 & $-$0.6 & $-$2.6 & 13300 & 0.00035 & 12     &    0.0 & $-5.1$ & $+0.4$ 
\enddata 
\tablenotetext{a}{Name of transition that is optimized for the Cloudy model.}
\tablenotetext{b}{Heliocentric velocity.}
\tablenotetext{c}{Relative velocity from the system center whose
                  heliocentric velocity is 209.9~\kms.}
\tablenotetext{d}{Absorption optical depth at the line center of
                  \ion{H}{1}~21-cm, if column (1) is \ion{H}{1}.}
\tablenotetext{e}{Spin temperature measured from \ion{H}{1}~21-cm
                  emission line, if column (1) is \ion{H}{1}.}
\tablenotetext{f}{Best Voigt profile fit value of column density of
                  the optimized line in column (1).}
\tablenotetext{g}{Best Voigt profile fit value of Doppler parameter of
                  the optimized line in column (1).}
\tablenotetext{h}{Best model parameter of metallicity.}
\tablenotetext{i}{Best model parameter of ionization parameter.}
\tablenotetext{j}{Gas temperature from the best model.}
\tablenotetext{k}{Total hydrogen volume density per cubic
                  centimeter. This depends very strongly on the
                  radiation field's absolute strength (see \S~5.5).}
\tablenotetext{l}{Thickness of the absorber assuming a plane-parallel
                  structure. This depends very strongly on the
                  radiation field's absolute strength (see \S~5.5).}
\tablenotetext{m}{Nitrogen Deficiency, compared to the Solar abundance
                  pattern.}
\tablenotetext{n}{We adopt the same parameters as for the 236~\kms\
                  cloud, as an example of acceptable model.}
\tablenotetext{o}{This is an example of the best fit models that have
                  acceptable range of parameters, \logz\ $>$ $-$1.0
                  and $\log U$ $>$ $-$6.0.}
\tablenotetext{p}{We adopt the typical metallicity of the SMC, as an
                  example of acceptable models.}
\end{deluxetable}
\clearpage

\begin{deluxetable}{ccccccc}
\tablecaption{Summary of Incident Radiation Models\label{t4}}
\tablewidth{0pt}
\tablehead{
\colhead{Model}                         &
\colhead{EBR}                           &
\colhead{MW}                            &
\colhead{LMC}                           &
\colhead{$\log n_{\gamma}$(EBR)$^a$}    &
\colhead{$\log n_{\gamma}$(MW+LMC)$^b$} &
\colhead{$f$$^c$}                       \\
\colhead{}                          &
\colhead{}                          &
\colhead{}                          &
\colhead{}                          &
\colhead{(\cmmm)}                   &
\colhead{(\cmmm)}                   &
\colhead{}                          \\
\colhead{(1)}                       &
\colhead{(2)}                       &
\colhead{(3)}                       &
\colhead{(4)}                       &
\colhead{(5)}                       &
\colhead{(6)}                       &
\colhead{(7)}                       \\
}
\startdata
1 & Y &   &      & $-$6.06 & ...     & ...  \\
2 & Y & Y &      & $-$6.06 & $-$6.22 & 0.69 \\
3 & Y & Y & 15\% & $-$6.06 & $-$4.84 & 16.6 \\
4 & Y & Y & 30\% & $-$6.06 & $-$4.54 & 33.1 
\enddata 
\tablenotetext{a}{Volume density of hydrogen-ionizing photons from the
  extragalactic background radiation \citep{haa96,haa01}.}
\tablenotetext{b}{Volume density of hydrogen-ionizing photons from the
  MW and the LMC \citep{fox05a}.}
\tablenotetext{c}{Ratio of the ionizing photon number densities from
  the MW and/or the LMC, to that from the EBR,
  $n_{\gamma}$(MW+LMC)/$n_{\gamma}$(EBR).}
\end{deluxetable}

\begin{deluxetable}{lccll}
\tabletypesize{\scriptsize}
\tablecaption{Model Constraints for the Magellanic Bridge Systems \label{t5}}
\tablewidth{0pt}
\tablehead{
\colhead{Cloud}         &
\colhead{parameter}     &
\colhead{constraint}    &
\colhead{line}          &
\colhead{condition}     \\
\colhead{(1)}           &
\colhead{(2)}           &
\colhead{(3)}           &
\colhead{(4)}           &
\colhead{(5)}           
}
\startdata
\ion{H}{1}~($V_{\odot}$=207~\kms)  & $\log U$ & $>$ $-$6.0        & \ion{C}{1}                           & to avoid over-production  \\
                                   &          & $<$ $-$4.8        & \ion{C}{1}                           & to avoid under-production \\
                                   & $\log Z$ & $\geq$ $-$0.5     & \ion{S}{2}                           & to avoid under-production \\
                                   &          & $\leq$ $-$0.5$^a$ & \ion{Ni}{2}                          & to avoid over-production  \\
\ion{H}{1}~($V_{\odot}$=236~\kms)  & $\log U$ & $>$ $-$6.1        & \ion{C}{1}                           & to avoid over-production  \\
                                   &          & $<$ $-$5.0$^b$    & \ion{Si}{2}                          & to avoid under-production \\
                                   & $\log Z$ & $>$ $-$0.8        & \ion{Si}{2}                          & to avoid under-production \\
                                   &          & $<$ $-$0.6$^a$    & \ion{Ni}{2}                          & to avoid over-production  \\
\ion{H}{1}~($V_{\odot}$=176~\kms)  & $\log U$ & $>$ $-$6.0$^c$    & \ion{C}{1}                           & to avoid over-production  \\
                                   &          & $<$ $-$5.0$^c$    & \ion{Si}{2}, \ion{O}{1}              & to avoid under-production \\
                                   & $\log Z$ & $>$ $-$1.0        & \ion{Si}{2}                          & to avoid under-production \\
                                   &          & $<$ $-$0.7        & \ion{Mg}{1}, \ion{Ni}{2}, \ion{C}{1} & to avoid over-production  \\
\ion{Si}{4}~($V_{\odot}$=217~\kms) & $\log U$ & $>$ $-$2.7        & \ion{C}{4}                           & to avoid under-production \\
                                   &          & $<$ $-$2.4        & \ion{C}{4}                           & to avoid over-production  \\
                                   & $\log Z$ & $>$ $-$4.0$^d$    & \ion{H}{1}~21-cm                      & to avoid over-production  \\
\ion{O}{1}~($V_{\odot}$=161~\kms)  & $\log U$ & $>$ $-$6.0        & \ion{Mg}{1}                          & to avoid over-production  \\
                                   &          & $<$ $-$3.5        & \ion{Si}{4}                          & to avoid over-production  \\
                                   & $\log Z$ & $\geq$ $-$1.0$^d$ & \ion{H}{1}~21-cm                      & to avoid over-production  \\
\enddata 
\tablenotetext{a}{Because nickel is strongly depleted on to the dust,
  this upper limit on the metallicity can be softened once the depletion
  is considered.}
\tablenotetext{b}{This constraint is valid if we assume \logz\ = $-0.7$.}
\tablenotetext{c}{This constraint is valid if we assume $-1.0$ $<$
  \logz\ $<$ $-0.7$.}
\tablenotetext{d}{No upper limit can be placed because the
  corresponding \ion{H}{1}~21-cm line is not detected with $>1\sigma$
  detection limit.}
\end{deluxetable}
\clearpage

\begin{deluxetable}{ccccccc}
\tabletypesize{\scriptsize}
\tablecaption{Comparison of Abundance Patterns$^a$ \label{t6}}
\tablewidth{0pt}
\tablehead{
\multicolumn{1}{c}{}              &
\multicolumn{5}{c}{$\log(element/H)$} &
\multicolumn{1}{c}{}              \\
\cline{2-6}    \\
\colhead{}     &
\colhead{C}    &
\colhead{N}    &
\colhead{O}    &
\colhead{Si}   &
\colhead{S}    &
\colhead{ref.$^b$} \\
\colhead{(1)}  &
\colhead{(2)}  &
\colhead{(3)}  &
\colhead{(4)}  &
\colhead{(5)}  &
\colhead{(6)}  &
\colhead{(7)}  
}
\startdata
The Sun                 & $-$3.61$\pm$0.04   & $-$4.17$\pm$0.11   & $-$3.31$\pm$0.05   & $-$4.46$\pm$0.05   & $-$4.80$\pm$0.05   & 1 \\
SMC \ion{H}{2} region   & $-$4.47$\pm$0.06   & $-$5.41$\pm$0.08   & $-$3.95$\pm$0.08   & $-$5.30$\pm$0.2    & $-$5.58$\pm$0.11   & 2 \\
                        & $-$4.84$\pm$0.04   & $-$5.54$\pm$0.12   & $-$3.98$\pm$0.08   & ...                & $-$5.51$\pm$0.14   & 3 \\
SMC star (AV~304)       & $-$5.15            & $-$5.11$\pm$0.17   & $-$3.90$\pm$0.16   & $-$5.27$\pm$0.02   & ...                & 4 \\
Bridge star (DI~1162)   & $-$4.94$\pm$0.33   & $-$5.30            & $-$4.00$\pm$0.13   & $-$5.19$\pm$0.06   & ...                & 4 \\
Bridge star (DGIK~975)  & $-$5.32$\pm$0.25   & $-$5.26            & $-$3.96$\pm$0.21   & $-$5.78$\pm$0.15   & ...                & 4 \\
Bridge gas ($V_{\odot}$ = 176~\kms) & $-$4.61 to $-$4.31 & $-$6.37 to $-$5.37 & $-$4.31 to $-$4.01    & $-$5.46 to $-$5.16 & $-$5.74 to $-$5.44 & 5 \\
Bridge gas ($V_{\odot}$ = 207~\kms) & $-$4.11            & $-$5.37 to $-$5.17 & $-$3.81               & $-$4.96            & $-$5.24            & 5 \\
Bridge gas ($V_{\odot}$ = 236~\kms) & $-$4.41 to $-$4.21 & $-$6.47 to $-$5.87 & $-$4.11 to $-$3.91    & $-$5.26 to $-$5.06 & $-$5.54 to $-$5.34 & 5 \\
Bridge gas (toward DI~1388)         & ...                & $-$5.29$\pm$0.11   & $-$4.30$+$0.13$-$0.11 & $-$5.45$+$0.14$-$0.12$^c$ & ...         & 6 \\
\enddata 
\tablenotetext{a}{All abundance patterns are measured from the
  observed spectra directly, except for our results for which we
  estimate them based on the photoionization model.}
\tablenotetext{b}{1: \citet{lod03}, 2: \citet{kur99}, 3:
  \citet{duf84}, 4: \citet{rol99}, 5: this paper, 6: \citet{leh08}}
\tablenotetext{c}{converted from $\log$(O/H), assuming solar abundance
  pattern.}
\end{deluxetable}
\clearpage


\begin{figure}
 \begin{center}
  \includegraphics[width=12cm,angle=0]{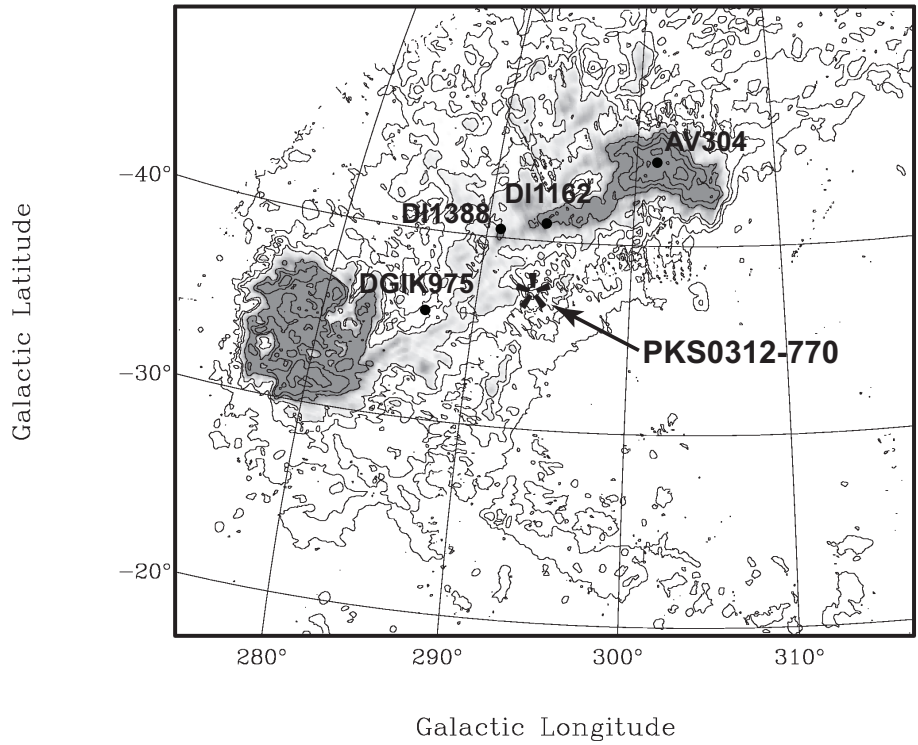}
 \end{center}
 \caption{Location of our target, PKS~0312$-$770, as well as stars in
   the MB (DGIK~975 and DI~1162, and DI~1388) and the SMC (AV~304),
   superimposed on the \ion{H}{1} peak brightness temperature ($T_B$)
   from \citet{put98}. Contours mark the 0.1~K (5~$\sigma$), 0.8, 2,
   8, 16, 32, 64, and 128~K peak brightness.  Light and deep gray
   contours correspond to $T_B$ $>$2~K and $>$8~K, respectively. Light
   gray contour roughly corresponds to the minimum threshold DLA
   column density, $N_{HI}$ = $10^{20.3}$~\cmm\ if the total line
   width of ion{H}{1}~21-cm absorption is similar to that toward
   PKS~0312$-$770, $dv$ $\sim$ 50~\kms\ (see Figure~5).}
\end{figure}

\begin{figure}
 \begin{center}
  \includegraphics[width=15cm,angle=0]{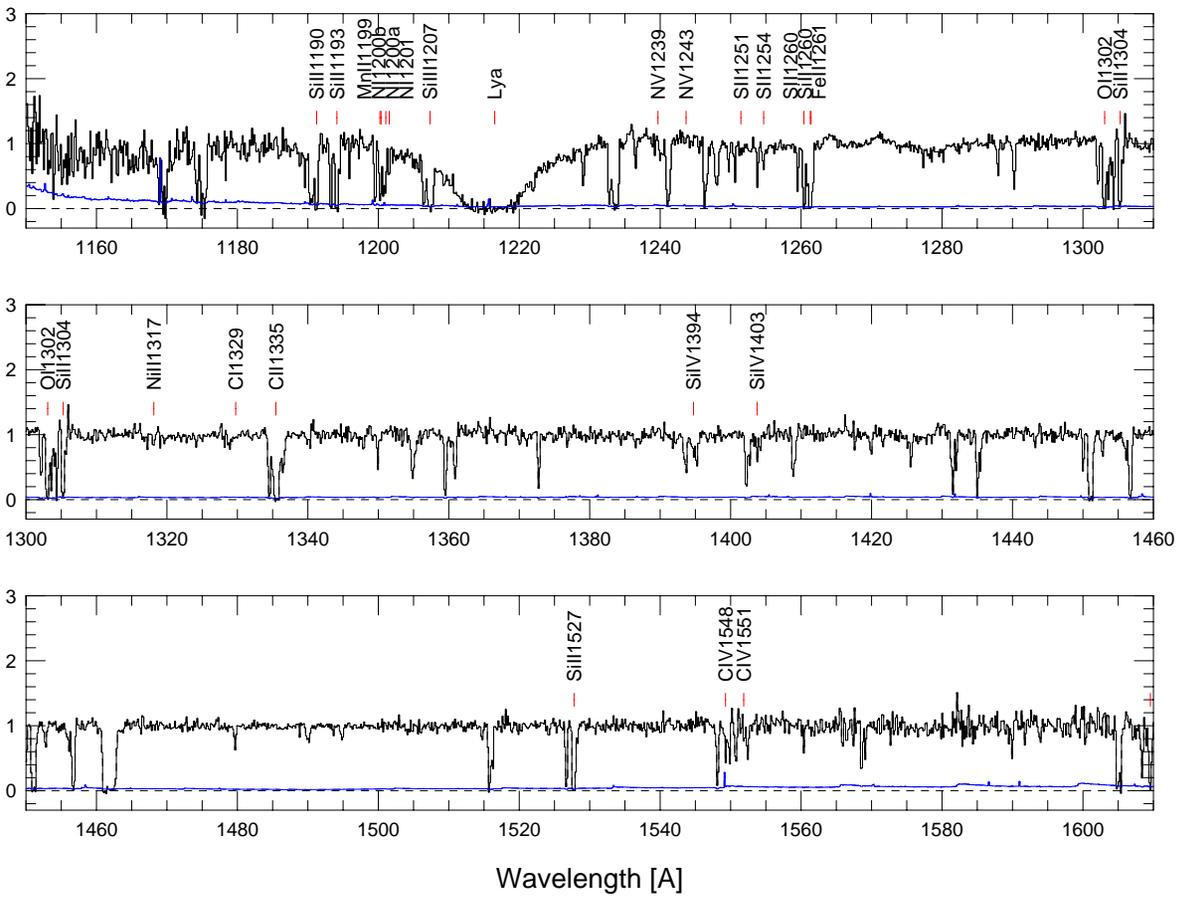}
 \end{center}
 \caption{Normalized flux versus wavelength for the {\it HST}/STIS
 E140M and E230M spectra of the quasar PKS~0312$-$770. The blue
 histogram displayed beneath the data represents the error
 spectrum. Positions of absorption lines through the MB at $z$
 $\sim$0.0007 toward the quasar are marked with ticks and transition
 names. The E140M spectrum at 1150\AA\ -- 1730\AA\ is binned every
 0.15\AA, while the E230M spectrum at 2130\AA\ -- 2980\AA\ is binned
 every 0.4\AA. A \lya\ emission around 1216\AA\ is removed.}
\end{figure}

\addtocounter{figure}{-1}
\begin{figure}
 \begin{center}
  \includegraphics[width=15cm,angle=0]{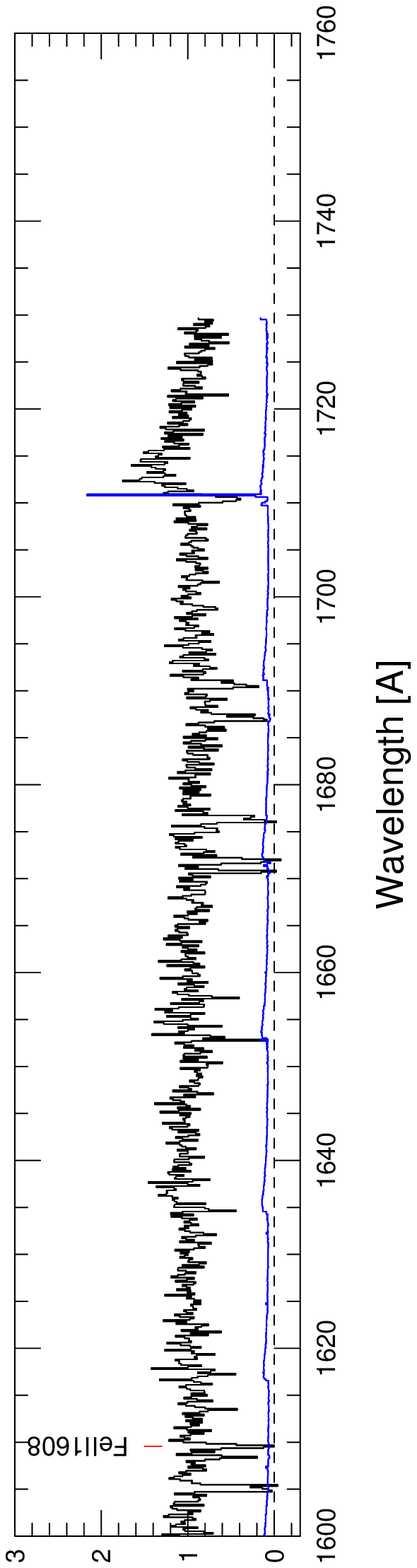}
 \end{center}
 \caption{Continued.}
\end{figure}

\addtocounter{figure}{-1}
\begin{figure}
 \begin{center}
  \includegraphics[width=15cm,angle=0]{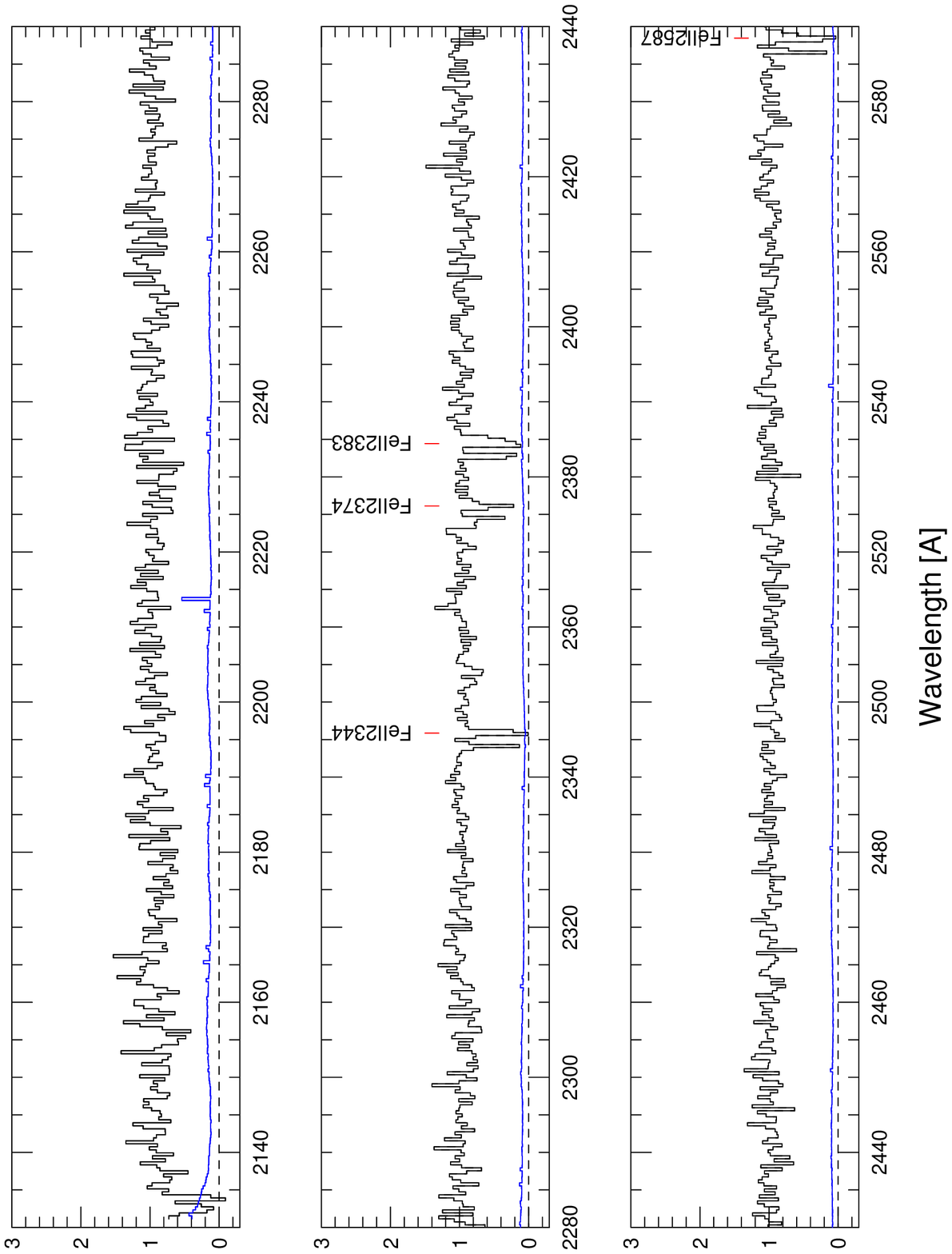}
 \end{center}
 \caption{Continued.}
\end{figure}

\addtocounter{figure}{-1}
\begin{figure}
 \begin{center}
  \includegraphics[width=15cm,angle=0]{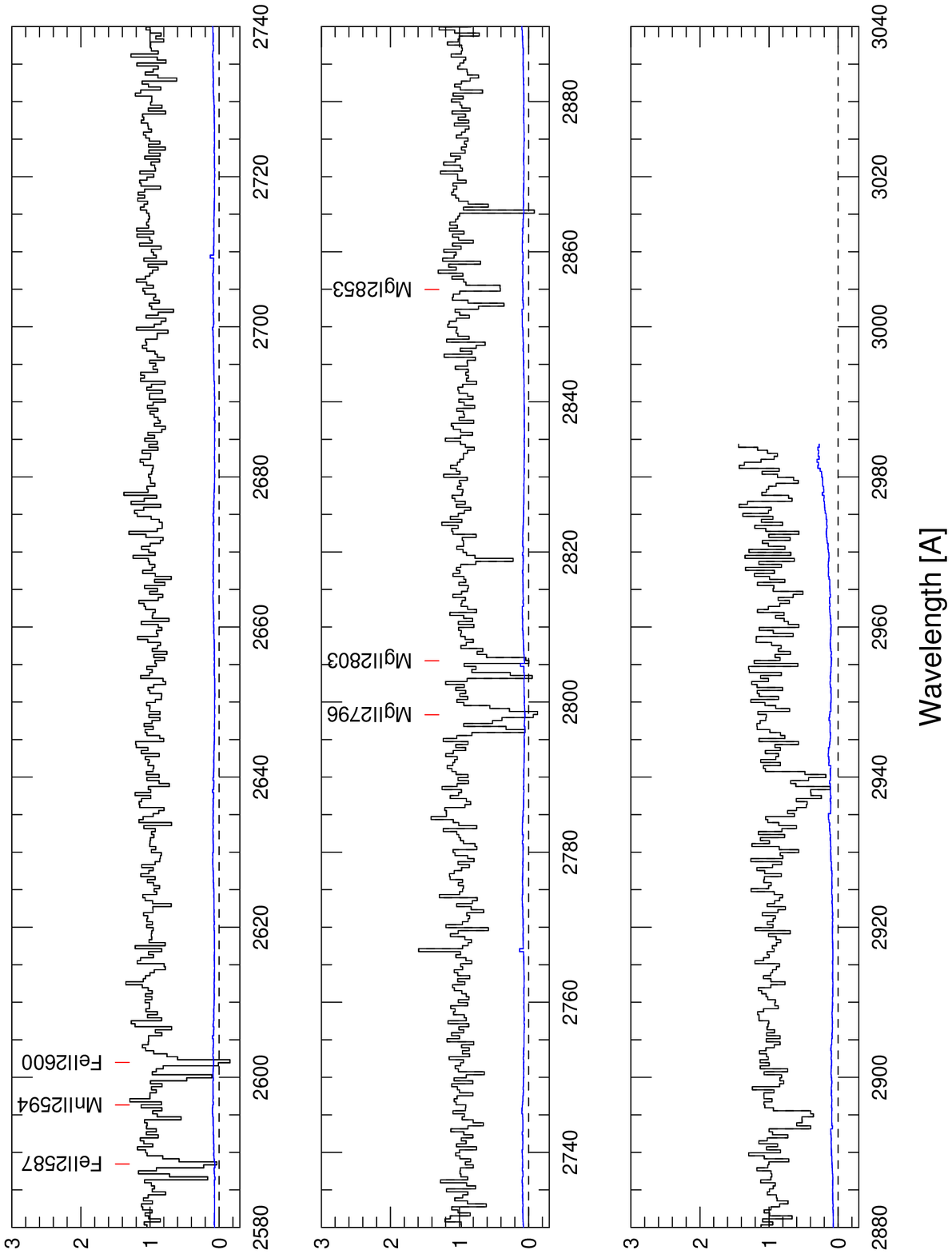}
 \end{center}
 \caption{Continued.}
\end{figure}

\begin{figure}
 \begin{center}
  \includegraphics[width=12cm,angle=0]{f3.eps}
 \end{center}
 \caption{Modeled spectrum with three \ion{O}{1}~$\lambda$1302
 components whose Doppler parameters are $b$ = 5.8, 7.6, and 6.3~\kms,
 from left to right \citep{kob99}. Positions of the lines are marked
 with ticks. Even if we synthesize spectra varying the column density
 of {\it each} component in the range \logn~(\ion{O}{1}/[\cmm]) = 15
 -- 19 (i.e., a total column density of all three components is in the
 range \logn~(\ion{O}{1}/[\cmm]) $\sim$15.5 -- 19.5), both sides of
 the observed spectrum shape cannot be reproduced.  We test a
 ridiculously wide range of column density by using same values for
 three \ion{O}{1} components without any justifications, because the
 purpose of this test is only examine whether the given Doppler
 parameters could reproduce both sides of the observed absorption
 profile.}
\end{figure}

\begin{figure}
 \begin{center}
  \includegraphics[width=12cm,angle=0]{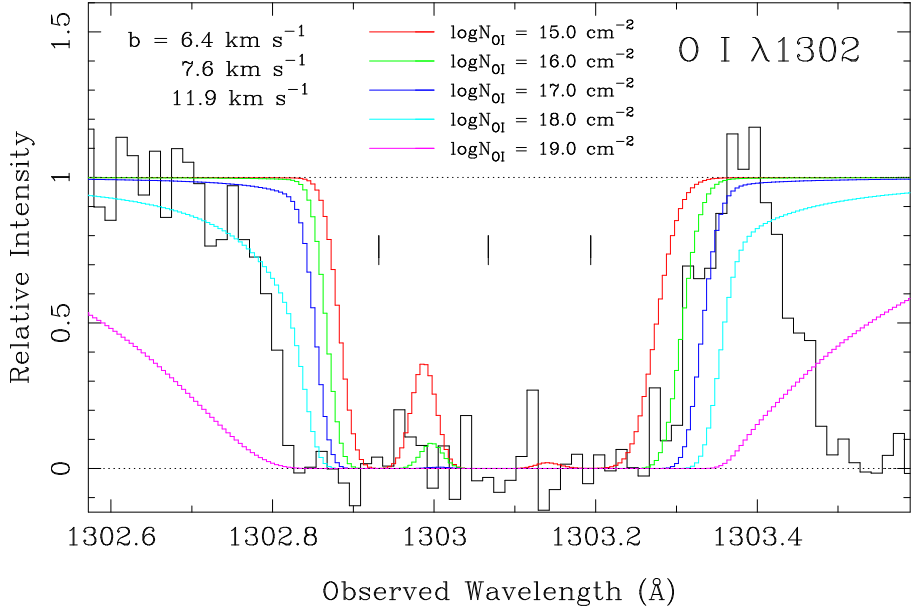}
 \end{center}
 \caption{Same as Figure~3, but the three \ion{O}{1}~$\lambda$1302
 components have Doppler parameters of $b$ = 6.4, 7.6, and 11.9~\kms\
 from our own fitting trials. A model with \logn~(\ion{O}{1}/[\cmm])
 of 16.0 gives an acceptable fit at the high-velocity side of the
 observed spectrum.}
\end{figure}

\begin{figure}
 \begin{center}
  \includegraphics[width=15cm,angle=0]{f5.eps}
 \end{center}
 \caption{Modeled spectrum of the \ion{H}{1}~21-cm absorption lines,
 overlaid with the radio spectrum of PKS~0312$-$770 taken with the
 Australia Telescope Compact Array \citep{kob99}. In addition to the
 original three components \citep{kob99}, we add another component at
 \delv\ = $-$49~\kms\ (i.e., $V_{\odot}$ = 160.8~\kms), required by
 the modeling of the metal lines. This additional component (whose
 central optical depth is $\tau$ = 0.022) is not detected at more than
 a 1$\sigma$ level in the observed spectrum whose S/N ratio is about
 40 per pixel (which corresponds to the detection limit of $\tau$ =
 0.025 at the line center).}
\end{figure}

\begin{figure}
 \begin{center}
  \includegraphics[width=10cm,angle=0]{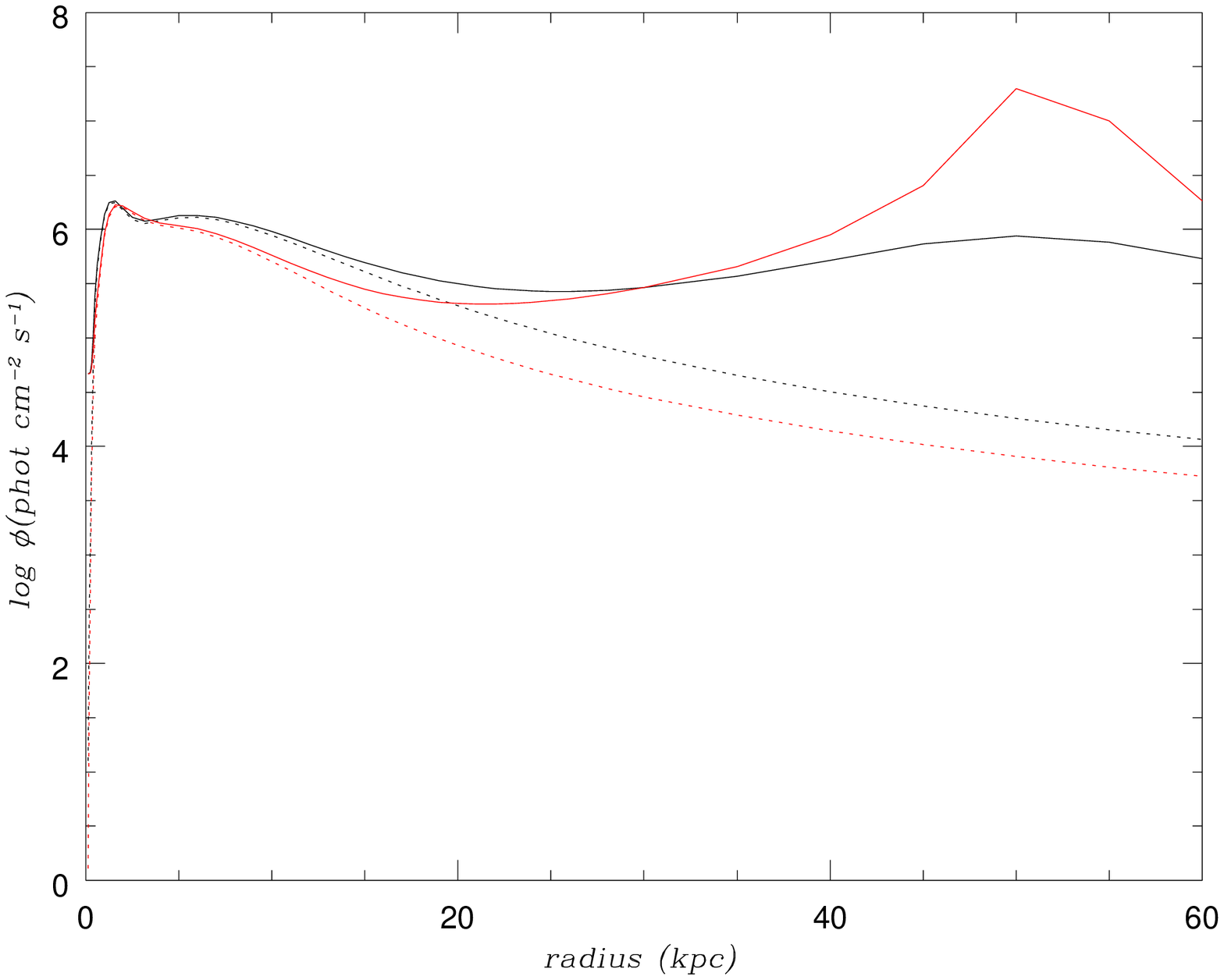}
 \end{center}
 \caption{Strength of the ionizing radiation field as a function of
 distance along our sight-line toward PKS~0312$-$770. The solid curve
 includes the radiation from the LMC as well as the radiation from the
 MW, while the dotted curve includes only the latter. The black curves
 are for the direction to PKS~0312$-$770, while the red curves are for
 the direction straight toward the LMC.  For example, at a radius of
 50~kpc, the ionization strength decreases from (i) LMC $+$ MW toward
 LMC, (ii) LMC $+$ MW toward PKS~0312$-$770, (iii) MW toward
 PKS~0312$-$770, to (iv) MW toward LMC.  The LMC model assumes that
 30~\% of the ionizing radiation escapes from the LMC, which we
 consider to be an upper limit.}
\end{figure}

\begin{figure}
 \begin{center}
  \includegraphics[width=9cm,angle=270]{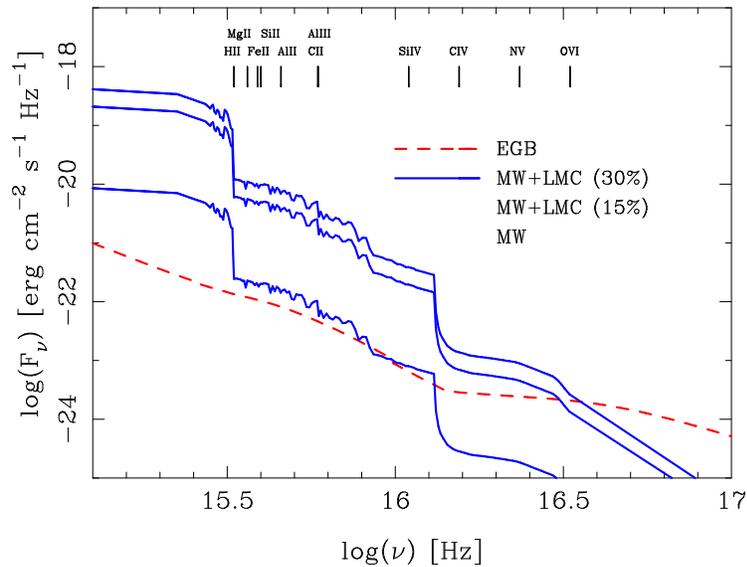}
 \end{center}
 \caption{Spectrum of four different radiation fields that we applied
 in our photoionization calculations with Cloudy. The smooth, dashed,
 red line denotes the extragalactic background radiation (EBR) from
 \citet{haa96,haa01}.  The three solid blue lines represent the
 radiation fields from the MW, MW $+$ LMC with 15\%\ escape fraction,
 and MW $+$ 30\%\ escape fraction, from bottom to top.  These are
 normalized assuming a distance of $D$ = 50~kpc from the center of the
 MW to the PKS~0312$-$770 sightline cloud. Ionization edges of several
 important transitions are indicated at the top of the plot.}
\end{figure}

\begin{figure}
 \begin{center}
  \includegraphics[width=12cm,angle=0]{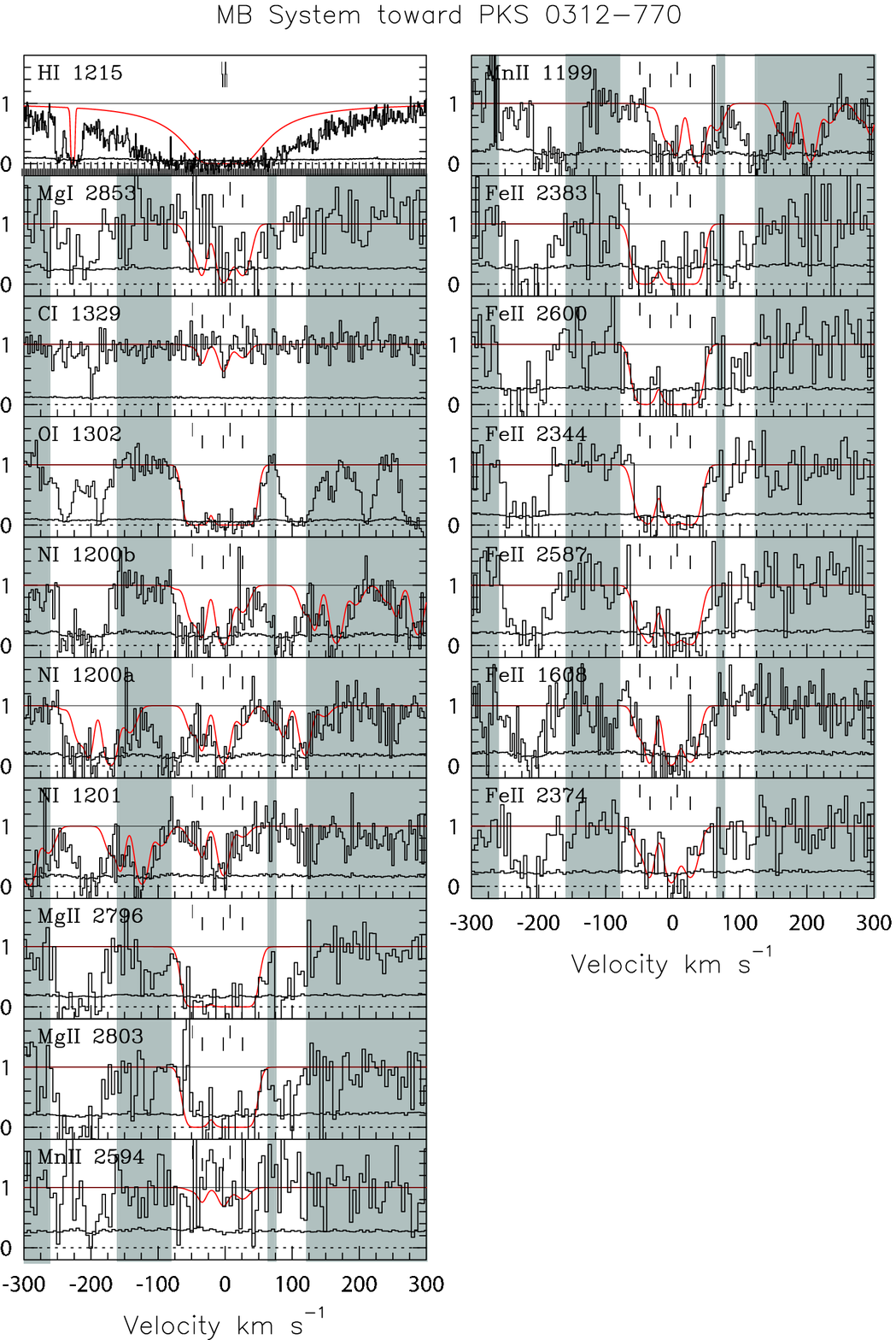}
 \end{center}
 \caption{Detected transitions, and those that provide limiting
 constraints, shown in velocity space for the absorption system in the
 Magellanic Bridge toward the quasar, PKS~0312-770. The velocity range
 is $\pm$300~\kms\ for all transitions except for \lya\ whose range is
 $\pm$3000~\kms. Velocity = 0~\kms\ corresponds to the system center
 of the Magellanic Bridge absorber whose heliocentric velocity is
 $V_{\odot}$ = 209.9~\kms. The data are from a {\it HST}/STIS
 spectrum. The error spectrum is also indicated as a solid histogram
 just above the dotted line crossing each plot at zero flux. An
 example of the best model fit to the observed spectrum (using model
 parameters listed in Table~3) is superimposed on the data as a solid
 (red) curve. Three unshaded regions denote the locations of the Milky
 Way, the MB, and a high-velocity cloud (from left to right),
 respectively.  The positions of absorption components are marked with
 ticks.  Above the spectrum in each panel, three ticks in the bottom
 line are \ion{H}{1} optimized components, while those in the top line
 are \ion{O}{1} or \ion{Si}{4} optimized components.}
\end{figure}
\clearpage

\addtocounter{figure}{-1}
\begin{figure}
 \begin{center}
  \includegraphics[width=12cm,angle=0]{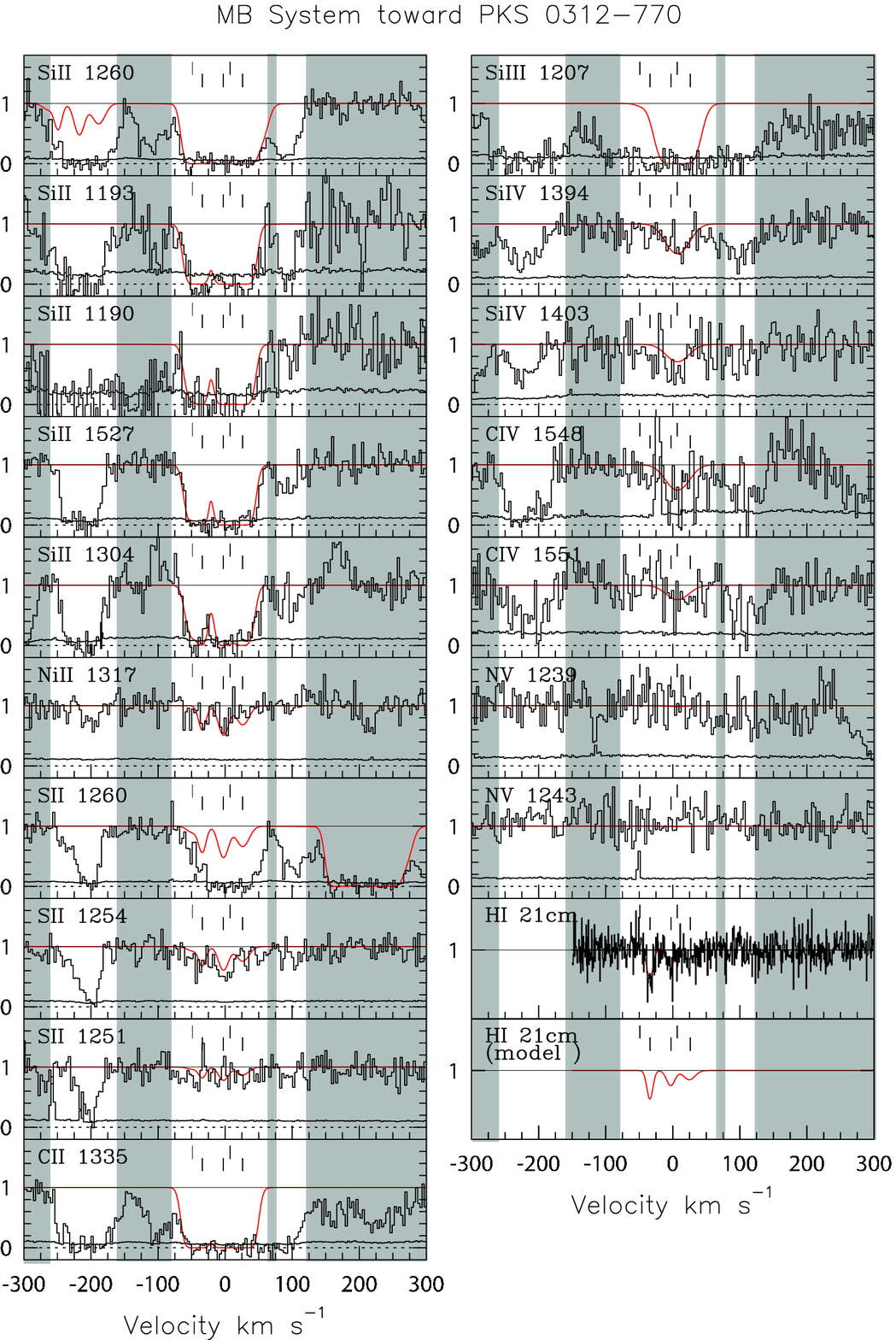}
 \end{center}
 \caption{Continued.}
\end{figure}
\clearpage

\begin{figure}
 \begin{center}
  \includegraphics[width=15cm,angle=270]{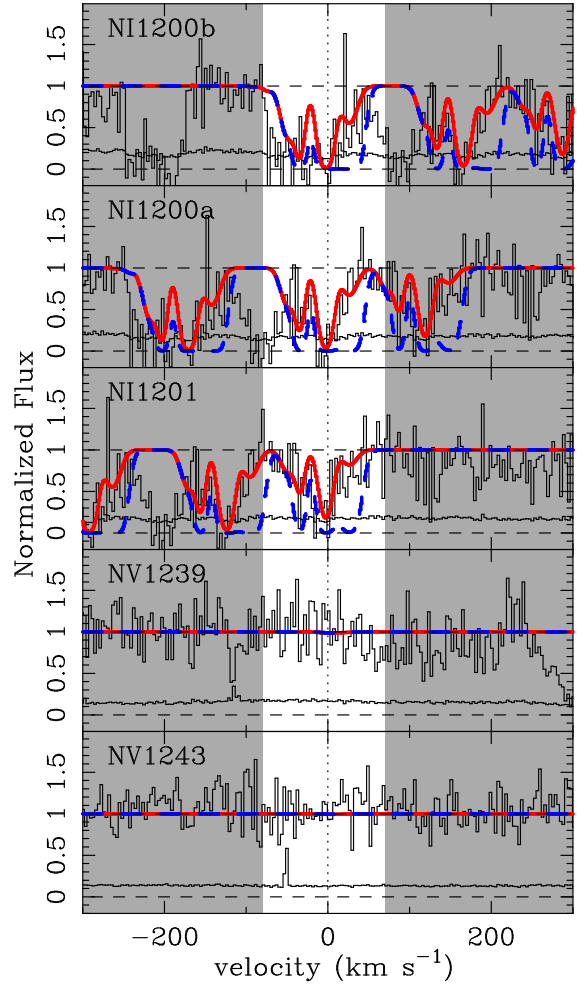}
 \end{center}
 \caption{Comparison of the model spectra with nitrogen deficiency
 (solid red line) and without nitrogen deficiency (dashed blue
 line). Obviously, the model with solar abundance pattern does not
 reproduce the observed \ion{N}{1} absorption lines. There is no large
 differences in the \ion{N}{5} doublet at this ionization parameter
 because of their very weak strengths. Unshaded parts are the regions
 of the absorber in the Magellanic Bridge.}
\end{figure}
\clearpage

\begin{figure}
 \begin{center}
  \includegraphics[width=15cm,angle=270]{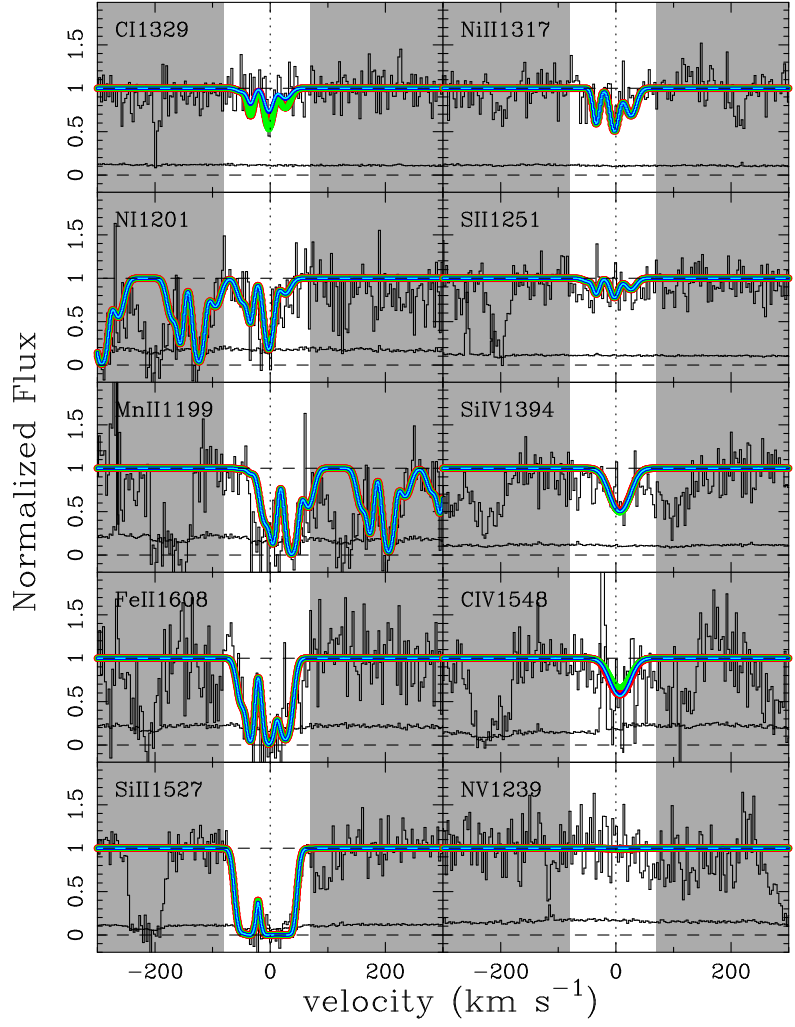}
 \end{center}
 \caption{Comparison of the model spectra using different incident
 radiation fields: (a) only extra-galactic background radiation (EGR)
 (red), (b) EGR plus radiation from the MW at a distance of $D$ =
 50~kpc from the MW (green), (c) EGB, MW radiation, plus radiation
 from the LMC with a 15~\% escape ratio as a minimum flux model (thick
 blue), and (d) EGB, the MW radiation, plus radiation from the LMC
 with a 30~\% escape ratio as a maximum flux model (thin blue).
 Unshaded parts are the regions of the absorber in the Magellanic
 Bridge. The color version of this figure is available in the
 electronic edition of the Journal.}
\end{figure}
\clearpage

\begin{figure}
 \begin{center}
  \includegraphics[width=12cm,angle=0]{f11.eps}
 \end{center}
 \caption{$\log$(N/Si) versus $\log$(Si/H) of three absorption
 components in the Magellanic Bridge toward PKS~0312$-$770, compared
 to those of damped Lyman alpha systems (dots; \citealt{hen07}), blue
 compact galaxies (blue cross; \citealt{nav06}) in which oxygen
 abundance was converted to silicon abundance using the solar
 abundance ratio \citep{hol01,all01}, and other SMC/MB objects from
 Table~6 (error bars).  Error bars shown in the bottom left are
 average uncertainties of the DLA systems in each direction.  The
 dotted error bar denotes the 1$\sigma$ error in our determination of
 the abundance ratio for the \ion{H}{1} component at $V_{\odot}$ =
 236~\kms, if we reduce the nitrogen abundance by $-$1.0~dex instead
 of $-$1.4~dex that is still consistent with the observed spectrum.
 The four possible evolutionary vectors are also shown in the bottom
 right. We do not plot components at $V_{\odot}$ = 161~\kms\ and
 217~\kms, because we cannot place any meaningful constraints on their
 abundance patterns.}
\end{figure}
\clearpage

\begin{figure}
 \begin{center}
  \includegraphics[width=15cm,angle=270]{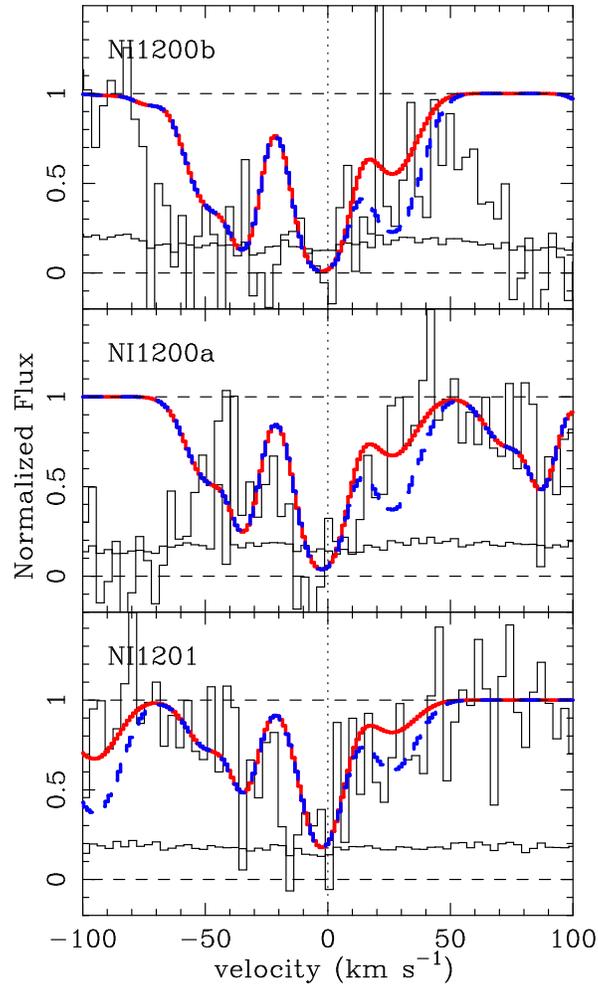}
 \end{center}
 \caption{Close up velocity plot around three detected \ion{N}{1}
   absorption lines within $\pm$100~\kms\ of the system center. Solid
   lines represent a model in which we decrease nitrogen abundance of
   the \ion{H}{1} component at $V_{\odot}$ = 237~\kms\ by 1.4~dex
   compare to the other elements (i.e., our best model in Table~3),
   while dashed lines are for the same model but with the nitrogen
   abundance reduced by 1.0~dex. If the continuum fitting around
   \ion{N}{1}~1200a is incorrect, this latter model would reproduce
   absorption features of the other two \ion{N}{1} lines much
   better. Moreover, this weaker nitrogen deficiency is more
   reasonable if we compare to those of the other two \ion{H}{1}
   components at $V_{\odot}$ = 176 and 207~\kms.}
\end{figure}

\end{document}